\documentclass[epj,nopacs,final]{svjour}
\usepackage{graphics}
\usepackage{latexsym}
\usepackage{subfigure,wrapfig,multirow}
\usepackage{epsfig,color,rotating,amsmath,delarray,array}
\usepackage{makeidx,pifont,float,amssymb}
\definecolor{gray01}{gray}{0.9}
\definecolor{gray02}{gray}{0.8}
\definecolor{gray03}{gray}{0.7}
\definecolor{gray04}{gray}{0.6}
\definecolor{gray05}{gray}{0.5}
\definecolor{gray06}{gray}{0.4}
\definecolor{gray07}{gray}{0.3}
\definecolor{gray08}{gray}{0.2}
\definecolor{gray09}{gray}{0.1}

\begin{document}

\title{Study of the reaction \boldmath$\gamma p\to p\pi^0\eta$}
\titlerunning{Study of the reaction $\gamma p\to p\pi^0\eta$}
\authorrunning{The CB-ELSA Collaboration}
\author{
The CB-ELSA Collaboration \medskip \\
I.~Horn$\,^1$, A.V.~Anisovich$\,^{1,2}$, G.~Anton$\,^3$,
R.~Bantes$\,^4$, O.~Bartholomy$\,^1$, R.~Beck$\,^1$,
Y.~Beloglazov$\,^2$, R.~Bogend\"orfer$\,^3$, R.~Castelijns$\,^5$,
V.~Cred\'e$\,^{6}$, A.~Ehmanns$\,^1$, J.~Ernst$\,^1$,
I.~Fabry$\,^1$, H.~Flemming$\,^{7,\rm a}$, A.~F\"osel$\,^3$,
M.~Fuchs$\,^1$, Ch.~Funke$\,^1$, R.~Gothe$\,^{4,\rm b}$,
A.~Gridnev$\,^{2}$, E.~Gutz$\,^1$, S.~H\"offgen$\,^4$, J.~H\"o\ss
l$\,^3$, J.~Junkersfeld$\,^1$, H.~Kalinowsky$\,^1$, F.~Klein$\,^4$,
E.~Klempt$\,^1$, H.~Koch$\,^7$, M.~Konrad$\,^4$, B.~Kopf$\,^7$,
B.~Krusche$\,^8$, J.~Langheinrich$\,^{4,\rm b}$, H.~L\"ohner$\,^5$,
I.~Lopatin$\,^2$, J.~Lotz$\,^1$, H.~Matth\"ay$\,^7$, D.~Menze$\,^4$,
J.~Messchendorp$\,^{9,\rm c}$, V.~Metag$\,^9$,
V.A.~Nikonov$\,^{1,2}$, D.~Novinski$\,^2$, M.~Ostrick$\,^{4,\rm d}$,
H.~van~Pee$\,^{1}$, A.~Radkov$\,^2$, A.V.~Sarantsev$\,^{1,2}$,
C.~Schmidt$\,^1$, H.~Schmieden$\,^4$, B.~Schoch$\,^4$,
G.~Suft$\,^3$, V.~Sumachev$\,^2$, T.~Szczepanek$\,^1$,
U.~Thoma$\,^{1}$, D.~Walther$\,^{1}$,~and Ch.~Weinheimer$\,^{1,\rm
e}$ }

\institute{$^1\,$Helmholtz-Institut f\"ur Strahlen- und Kernphysik,
Universit\"at Bonn, Germany\\
$^2\,$Petersburg Nuclear Physics Institute, Gatchina, Russia\\
$^3\,$Physikalisches Institut, Universit\"at Erlangen, Germany\\
$^4\,$Physikalisches Institut, Universit\"at Bonn, Germany\\
$^5\,$Kernfysisch Versneller Instituut,
Groningen, The Netherlands\\
$^6\,$Department of Physics, Florida State University, Tallahassee,
FL, USA\\
$^7\,$Institut f\"ur Experimentalphysik I, Universit\"at Bochum,
Germany\\
$^8\,$Institut f\"ur Physik, Universit\"at Basel,
Switzerland\\
$^9\,$II. Physikalisches Institut, Universit\"at
Gie{\ss}en, Germany\\[1mm]
$^{\rm a}\,$Present address: GSI, Darmstadt, Germany\\
$^{\rm b}\,$Present address:
University of South Carolina, Columbia, SC, USA\\
$^{\rm c}\,$Present
address: Kernfysisch Versneller Instituut, Groningen, The Netherlands\\
$^{\rm d}\,$Present address: Institut f\"ur Kernphysik, Universit\"at
Mainz, Germany\\
$^{\rm e}\,$Present address: Institut f\"ur
Kernphysik, Universit\"at M\"unster, Germany\\
}

\abstract{The reaction $\gamma p\to p\pi^0\eta$ has been studied
with the CBELSA detector at the tagged photon beam of the Bonn
electron stretcher facility. The reaction shows contributions from
$\Delta^+(1232)\eta$, $N(1535)^+\pi^0$ and $pa_0(980)$ as
intermediate states. A partial wave analysis suggests that the
reaction proceeds via formation of six $\Delta$ resonances,
$\Delta(1600)P_{33}$, $\Delta(1920)P_{33}$, $\Delta(1700)D_{33}$,
$\Delta(1940)D_{33}$, $\Delta(1905)F_{35}$, $\Delta(2360)D_{33}$,
and two nucleon resonances $N(1880)P_{11}$ and $N(2200)P_{13}$, for
which pole positions and decay branching ratios are given.
 \vspace*{1mm} \\ {\it PACS: 13.30.-a Decays of baryons,  13.60.Le
Meson production
  14.20.Gk Baryon resonances with $S=0$}
}
\date{Received: \today / Revised version:}

\mail{klempt@hiskp.uni-bonn.de}

\maketitle

\section{Introduction}
The study of baryon resonances has found renewed interest as
evidenced by new experimental facilities providing continuous photon
beams like Jefferson Lab, SPring-8, MAMI at Mainz, and ELSA at Bonn.
For nucleon resonances up to $\approx 1.7$\,GeV/c$^2$, their
abundance and most masses are reasonably well explained in quark
models even though the models use very different assumptions on the
nature of long-range quark-quark interactions, like effective gluon
exchange \cite{Capstick:1986bm}, instanton induced interactions
\cite{Loring:2001kx}, or exchange of Goldstone bosons \cite{Riska}.
At higher masses, the models differ in important details of their
predictions but agree in predicting the existence of many more
states than have been found experimentally. Diquark models
\cite{Anselmino:1992vg} have a reduced number of degrees of freedom
and expect fewer states \cite{Kirchbach:2001de,Santopinto:2004hw}.
Models based on a conformal approximation of QCD exploit the
correspondence \cite{Maldacena:1997re} between string theories on an
Anti-de-Sitter (AdS) space and gauge theories on its space-time
boundary \cite{Witten:1998qj,Klebanov:2000me}. Apparently, AdS/QCD
predicts a smaller number of states
\cite{Karch:2006pv,Brodsky:2006uq,Forkel:2007cm,Forkel:2007tz} but,
to our knowledge, the problem of {\it missing resonances}, of
resonances predicted in quark models but not in AdS/QCD, has not
\begin{figure*}[pt]
\vspace{4mm}
\begin{center}
\epsfig{file=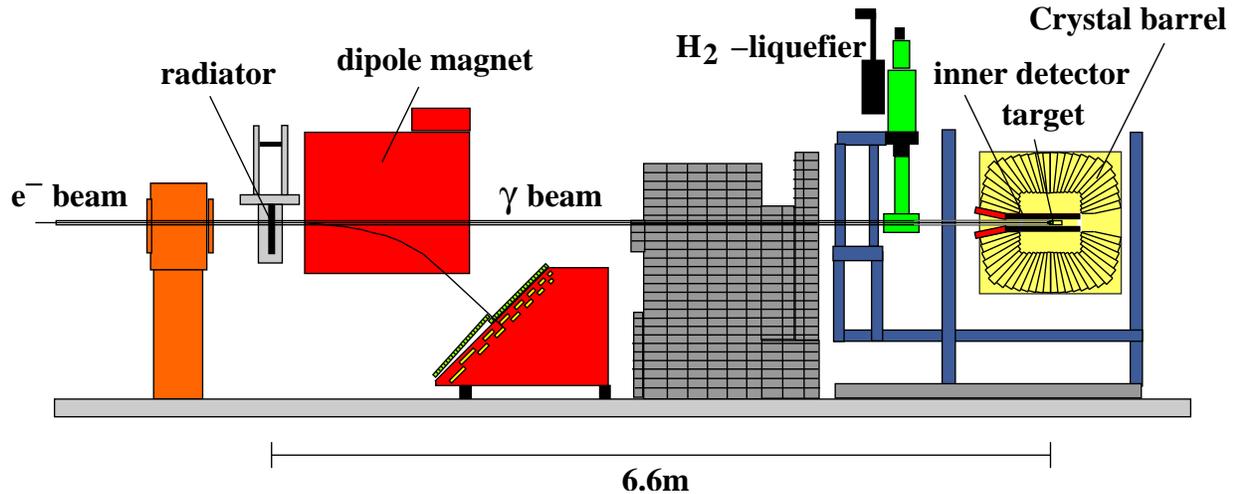,width=.90\textwidth}
\end{center}
\caption{Experimental setup at ELSA in Bonn:  Electrons delivered by
the accelerator ELSA enter the experimental area from the left. They
hit a radiator and produce a wide-band photon beam. The photon
energy is tagged by measuring the electron deflecting angle in a
dipole magnet. The photons hit protons of a liquid H$_2$ target.
Particles, in particular photons, emerging from the reaction are
detected in the Crystal barrel detector. A scintillation fiber
(scifi) detector measures the direction of charged particles. A
total absorption lead mineral oil $\check{\rm C}$erenkov counter
installed further downstream to monitor the photon beam intensity is
not shown in the figure.}
\label{FigureCBELSAExperiment}
\end{figure*}
yet been discussed. Lattice gauge calculations aim at simulating
full QCD; substantial progress has been achieved
\cite{Basak:2007kj}, however, the calculations do not yet help to
validate or reject specific quark models.

In the past, knowledge on nucleon resonances was derived mostly from
partial wave analyses of elastic $\pi N$ scattering experiments.
This method exploits the coupling of resonances to the $N\pi$
channel; resonances for which this coupling is weak are not
detectable. The Particle Data Listings \cite{Yao:2006px} teach us
that at high pion momenta, above 2\,GeV, high-spin states are formed
while formation of high-mass states with low angular momenta seems
to be suppressed. Due to this limitation, only few resonances are
known so far. Even worse, the evidence for a large fraction of them
has recently been questioned in a careful analysis
\cite{Arndt:2006bf} of a large body of $N\pi$ elastic and
charge-exchange scattering data.

The new facilities give access to photo-induced production of
resonances; detectors with large solid angle allow one to study
decays of resonances into complex final states. Thus $N\pi$ is
avoided in the entrance and in the exit channel. Quark model
calculations \cite{Metschpr} and first results
\cite{Anisovich:2005tf,Sarantsev:2005tg} seem to support our
conjecture that in photoproduction, excitation of radially excited
states might be preferred over excitations of high-angular momentum
states.

In this paper we present a study of the reaction
\begin{equation} \label{R1} \gamma p \to  p\pi^0\eta\ \end{equation}
for photon energies covering the resonance region from the
$p\pi^0\eta$ production threshold at $W=1.63$\,GeV/c$^2$ up to 2.5
GeV/c$^2$. In this reaction, decays of $\Delta$ resonances into
$\Delta(1232)\eta$ can be studied: the $\eta$ acts as an isospin
filter and resonances decaying into $\Delta(1232)\eta$ must have
isospin $I=3/2$. For low photon energies phase space is limited, and
$\Delta(1232)$ and $\eta$ should be in a relative S-wave. We thus
may expect a high sensitivity for baryon resonances with isospin
$I=3/2$ and spin $J=3/2$ and negative parity. If such resonances
decay into $N\pi$, they need $L=2$ between $N$ and $\pi$; resonances
with these quantum numbers are characterized by $L_{2I,2J}=D_{33}$.
The lowest mass resonance with these quantum numbers is
$\Delta(1700)D_{33}$. The possibility that it may couple to the
$\Delta(1232)\eta$ channel was already discussed by Nefkens
\cite{Nefkens:1997cc}; his conjecture was confirmed recently by the
GRAAL collaboration \cite{Ajaka:2008}. The aim of this paper is to
report evidence for $\Delta$ resonances above $\Delta(1700)D_{33}$,
in particular for a $J^P=3/2^{\pm}$ parity doublet consisting of the
two resonances $\Delta(1920)P_{33}$ and $\Delta(1940)D_{33}$.

\section{Experimental setup and data reconstruction}

The experiment was carried out at the tagged photon beam of the {\bf
EL}ectron {\bf S}tretcher {\bf A}ccelerator ELSA at Bonn
\cite{Hillert:2006yb}, using the Crystal Barrel detector
\cite{Aker:1992ny}. A short description of the experiment, data
reconstruction and analysis methods can be found in a recent letter
on $2\pi^0$  photoproduction \cite{Thoma:2007bm}; further details
are given in \cite{van Pee:2007tw}. A schematic drawing of the
experimental setup is shown in Fig.~\ref{FigureCBELSAExperiment}.
For the data presented here, ELSA delivered - via slow extraction -
a continuous electron beam of 3.2\,GeV. The electrons hit a radiator
target of 0.003\,$X_{\rm R}$ (i.e. radiation length) thickness.

\begin{figure*}[pt]
\centerline{
\epsfig{file=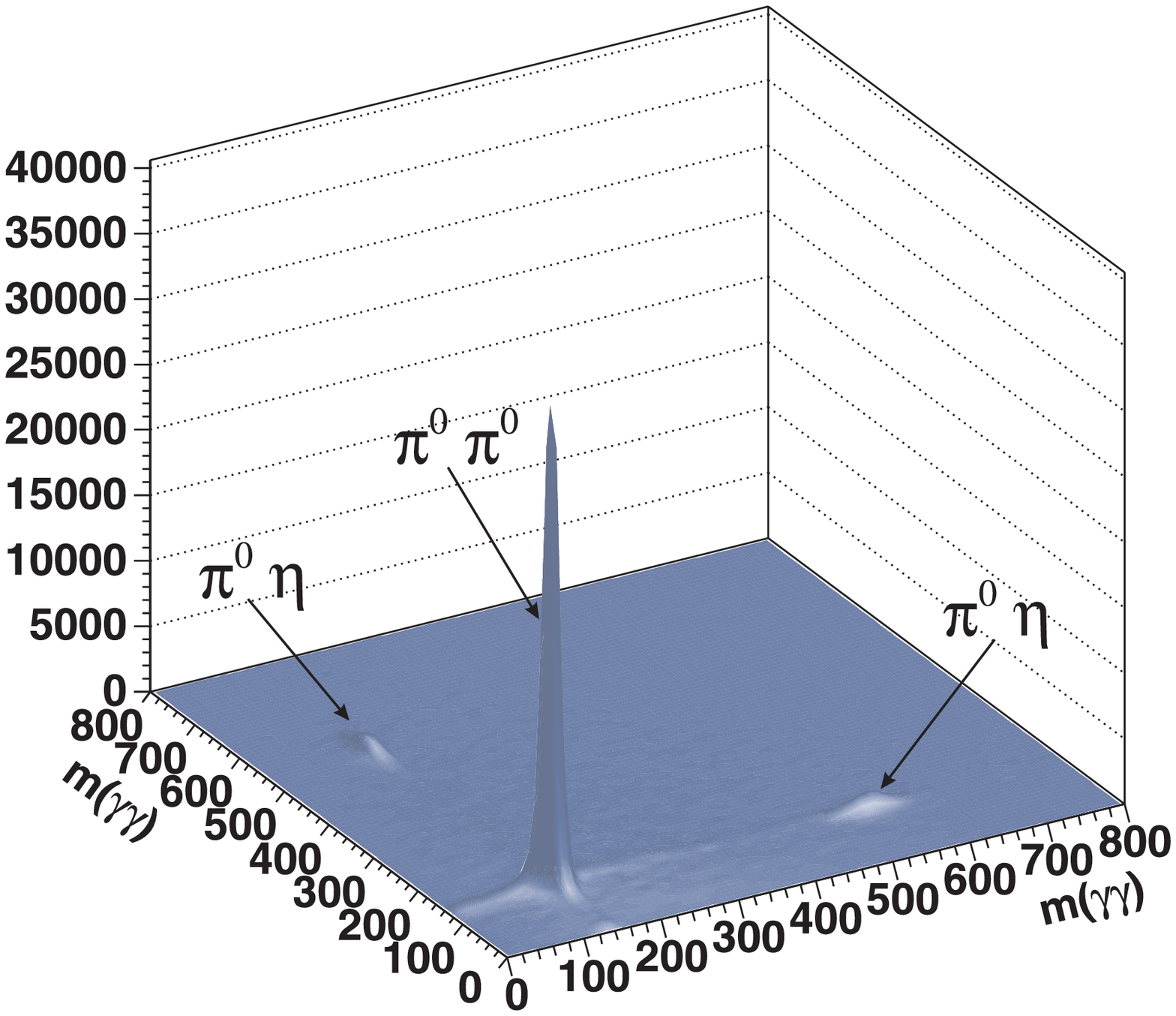,width=0.49\textwidth,height=0.35\textwidth,clip=}
\hspace{1mm} \epsfig{file=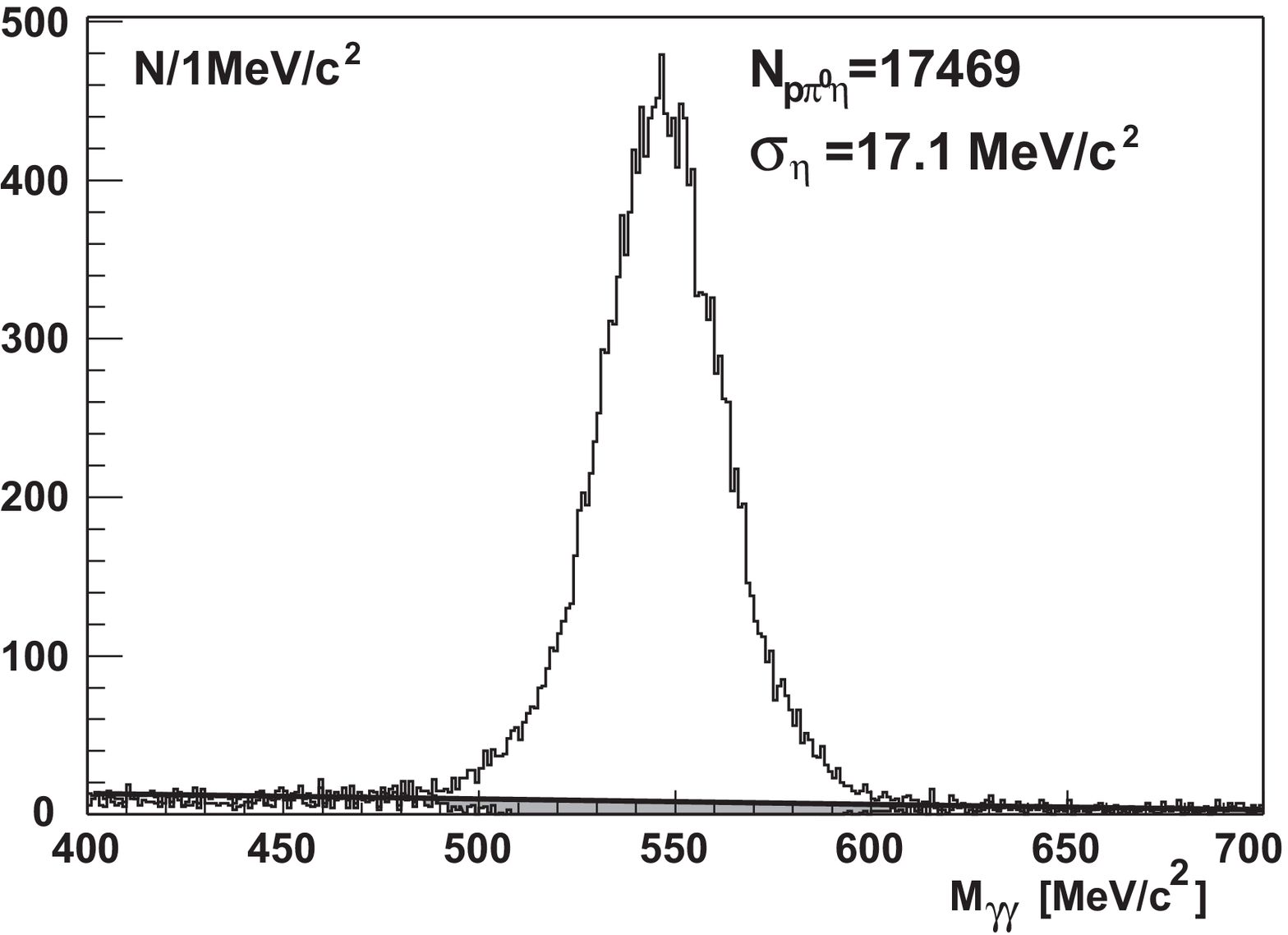,width=0.49\textwidth,clip=}}
\caption{\label{gg_vs_gg}Left: The distribution of $\gamma\gamma$
invariant masses after a kinematic fit to $\rm\gamma p \to p
4\gamma$ (with c.l.$>0.1$\%). There are six entries per event. The
distribution shows a large peak due to the reaction  $\rm\gamma p
\to p \pi^0\pi^0$ and two smaller peaks due to $\rm\gamma p \to p
\pi^0\eta$. Right: The $\gamma\gamma$ invariant mass distribution
after a kinematic fit to $\rm\gamma p \to p \pi^0\gamma\gamma$ (with
c.l.$>10$\%) and a cut rejecting $\rm\gamma p \to p \pi^0\pi^0$
events (with c.l.$>1$\%). In the 460-620\,MeV/c$^2$ mass interval
there are 18589 events, 1120 of them are below the background line
derived from a linear fit. The fit to $\rm\gamma p \to p \pi^0\eta$
(with c.l.$>1$\%) accepts 18379 events and rejects 210 background
events leading to 910 background events (grey-shaded area). }
\end{figure*}

The energies of photons produced in the radiator target were tagged
in energy by a measurement of the deflection angle of scattered
electrons passing the field of a dipole magnet. For singly scattered
electrons, the photon energy is $E_{\gamma}=E_0-E_{{\rm e}^-}$. The
tagging detector consisted of 2 Multi-Wire Proportional Chambers
(MWPCs) and 14 plastic scintillator bars. The position resolution of
the MWPCs determined the photon energy resolution of 0.5\,MeV at the
highest $E_{\gamma}$ and 30\,MeV at the lowest $E_{\gamma}$, whereas
the scintillation counters enabled fast timing. The energy
calibration of the tagger was performed by direct injection of the
electron beam.

The photon beam hit a liquid H$_2$ target (length: $l = 52.84$\,mm,
diameter: $d = 30$\,mm); charged reaction products were detected in
an inner three--layer scintillating--fiber detector surrounding the
target and positioned at mean radii of 5.81\,cm, 6.17\,cm and
6.45\,cm, respectively. The 2\,mm fibers are partly bent to helical
shapes ($-25^{\circ}$, $+25^{\circ}, 0^{\circ}$ from the inner to
the outer layer) to provide an unambiguous impact point when a
charged particle crosses the detector \cite{Suft:2005}.

The Crystal--Barrel calorimeter consisted of 1380
16-radiation-length CsI(Tl) crystals. The crystals are of
trapezoidal shape and point to the center of the target; they
provide an excellent photon detection efficiency and a high
granularity. The setup covered about 98\% of $4\pi$ (full $\phi$
coverage, $12^{\circ}\le\theta\le 168^{\circ}$). Its large
solid--angle coverage allowed for reconstruction of multi--photon
final states.

The first-level trigger was derived from a coincidence between the
tagging system and the fiber detector. In the second-level trigger,
a FAst Cluster Encoder (FACE) based on cellular logic, provided the
number of charged and neutral particles detected in the Crystal
Barrel. Data were taken triggering on events with two (partly three)
or more particles in the cluster logic. A segmented total-absorption
oil $\check{\rm C}$erenkov counter determined the total photon flux.

Events due to reaction (\ref{R1}) were selected by requiring five
clusters of energy deposits in the Crystal Barrel calorimeter. One
of the clusters was required to match with the charged particle
emerging from the liquid H$_2$ target and hitting the scintillation
fiber detector. The latter cluster was identified as proton, the
other four as photons. The proton and the four photons were assumed
to be produced in the target center. More details on event
reconstruction can be found in \cite{van Pee:2007tw}.

These events were subjected to a kinematic fit to the $\gamma p\to p
4\gamma$ hypothesis imposing energy and momentum conservation. The
distribution of the $\gamma\gamma$ invariant mass of one pair
against the second pair of surviving events (with 6 entries per
event) is shown in Fig. \ref{gg_vs_gg}, left panel. Then, the
$\gamma p\to p\,\pi^0\gamma\gamma$ hypothesis was tested and events
with a confidence level (c.l.) exceeding 10\% were retained. In a
next step, events compatible at a c.l.\,$>$1\% with the $\gamma p\to
p\,2\pi^0$ hypothesis were rejected. The resulting $\gamma\gamma$
invariant mass of the second photon-pair is shown in
Fig.~\ref{gg_vs_gg}, right panel. These events passed a final
kinematic fit to the $\gamma p\to p\,\pi^0\eta$ hypothesis requiring
a probability exceeding 1\%. The final event sample contains 17469
events due to reaction (\ref{R1}) and 910 background events.  Events
which are likely due to the background are identified (and
subtracted) by selecting those 910 events which are closest in phase
space to the events falling into the $\eta$ side bins
(380-440\,MeV/c$^2$; 640-700\,MeV/c$^2$).

\section{Mass and angular distributions}

\begin{figure*}[pt]
\begin{center}
\begin{tabular}{ccc}
\hspace{-2mm}\epsfig{file=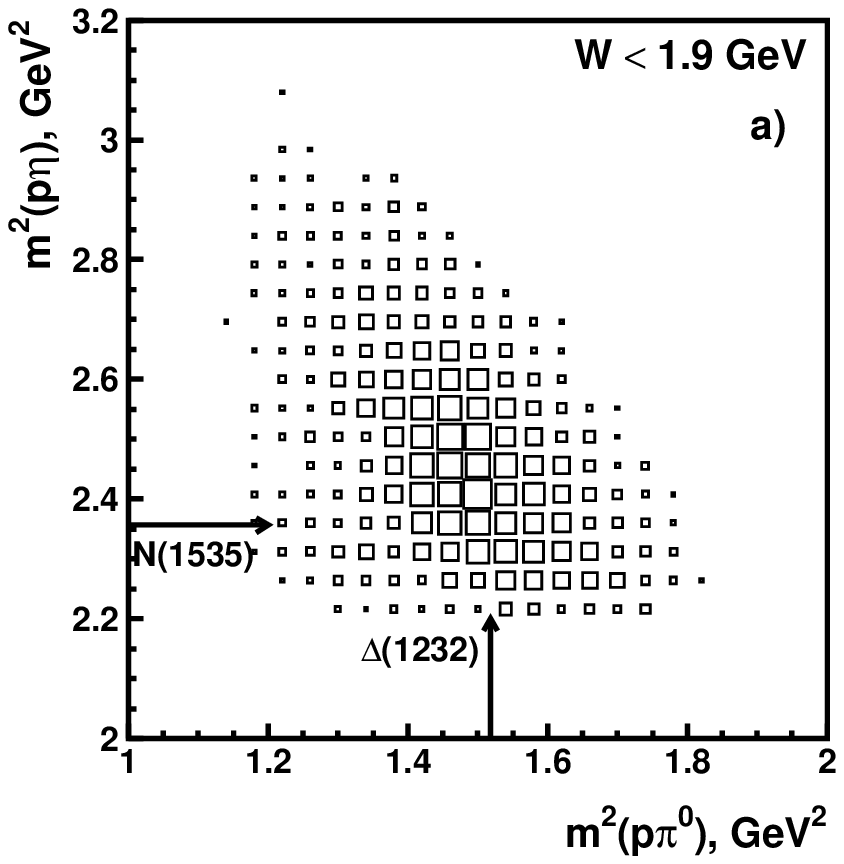,width=0.30\textwidth,clip=}&
\hspace{-6mm}\epsfig{file=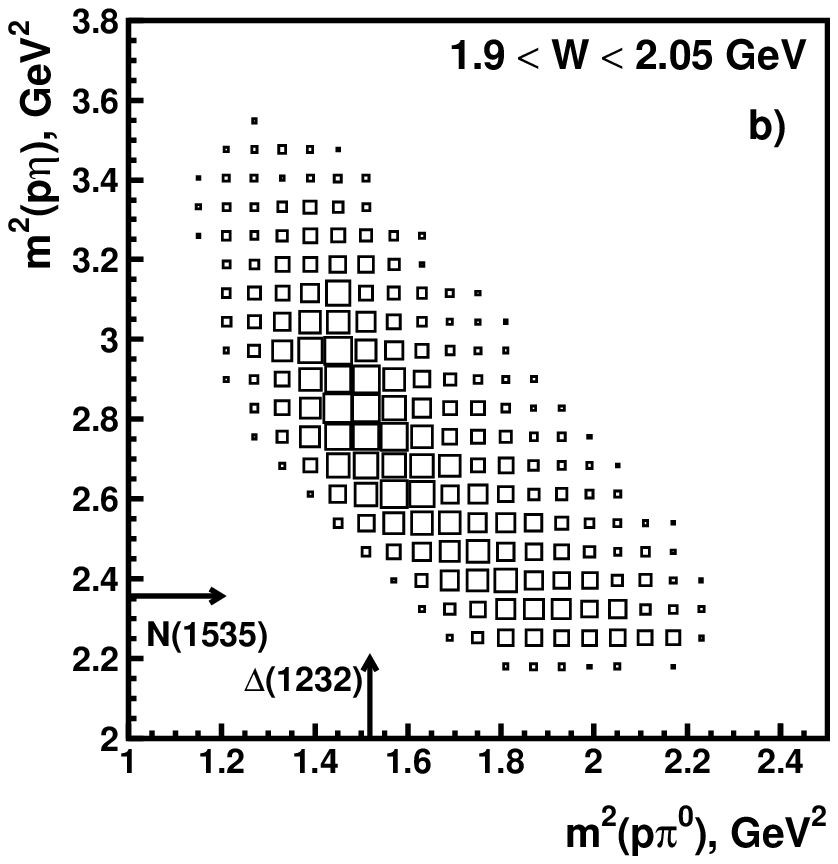,width=0.30\textwidth,clip=}&
\hspace{-6mm}\epsfig{file=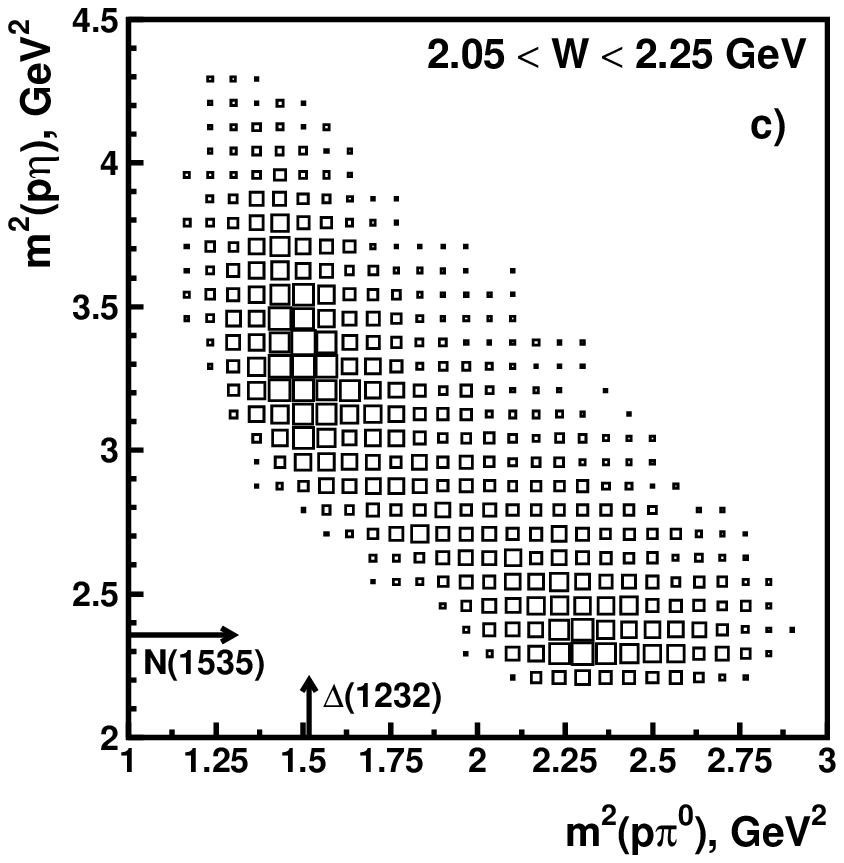,width=0.30\textwidth,clip=}\vspace{-3mm}\\
\hspace{-2mm}\epsfig{file=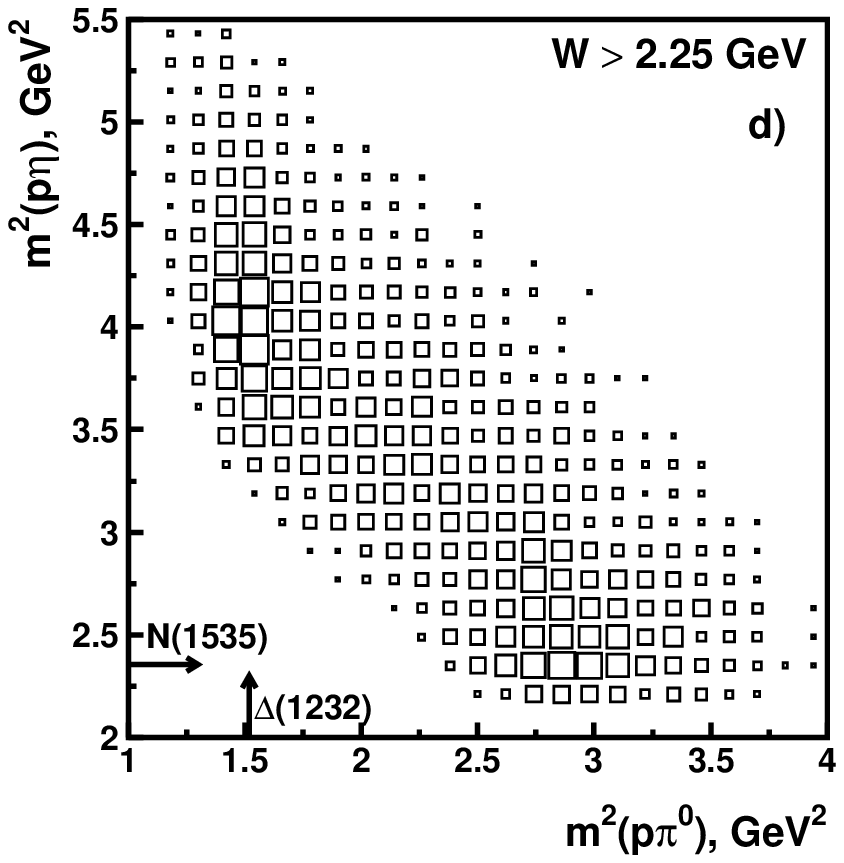,width=0.30\textwidth,clip=}&
\hspace{-6mm}\epsfig{file=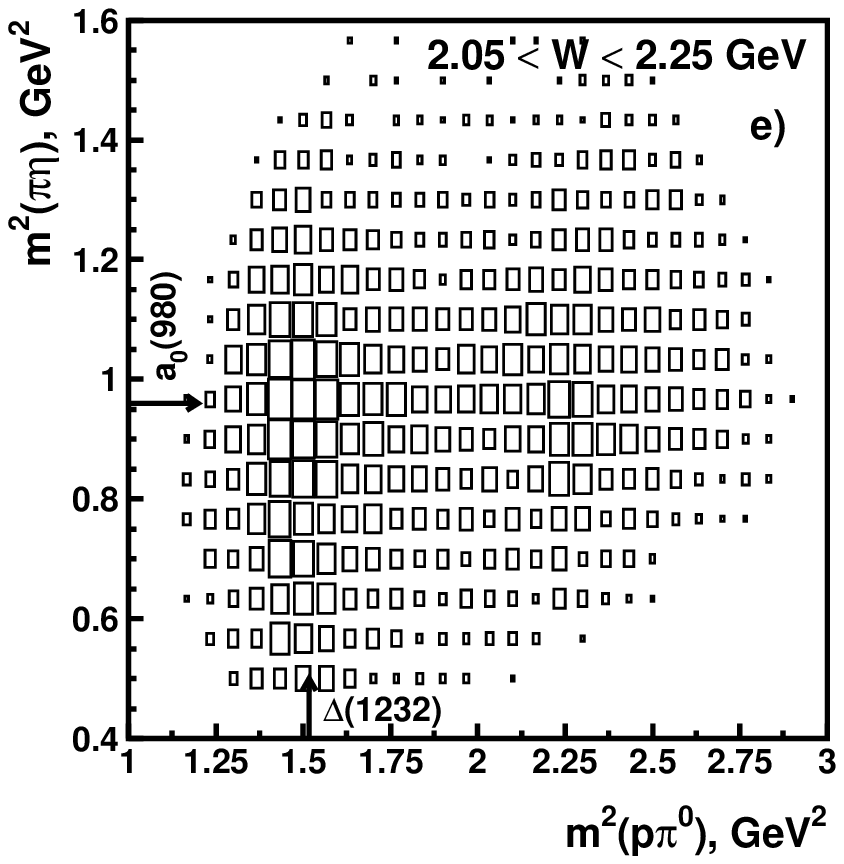,width=0.30\textwidth,clip=}&
\hspace{-6mm}\epsfig{file=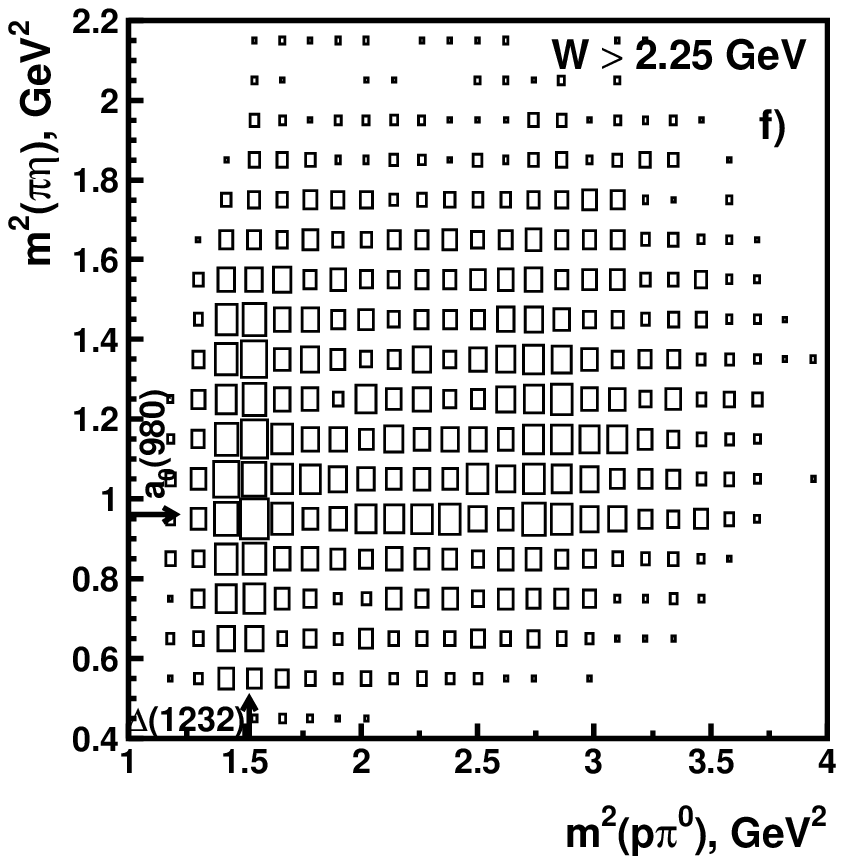,width=0.30\textwidth,clip=}\vspace{3mm} \\
\hspace{-2mm}\epsfig{file=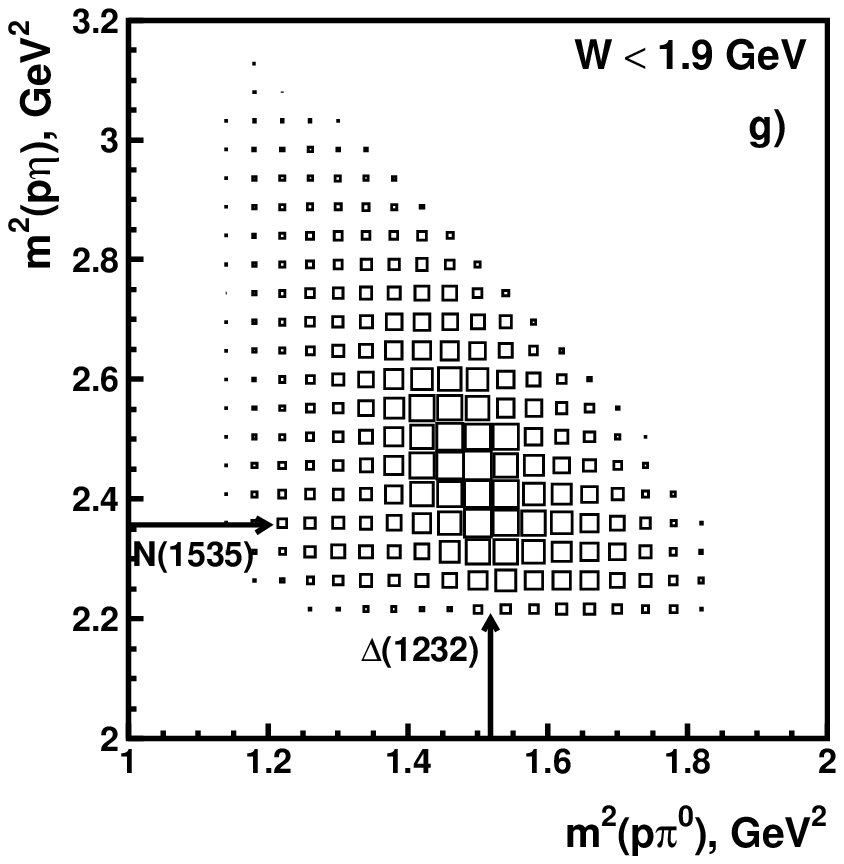,width=0.30\textwidth,clip=}&
\hspace{-6mm}\epsfig{file=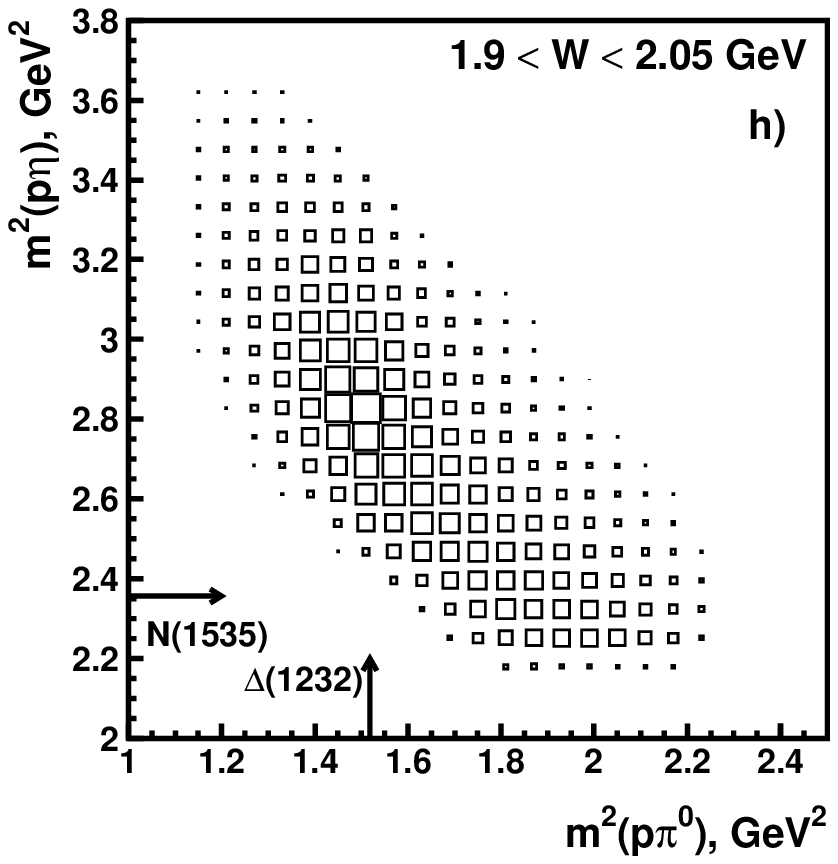,width=0.30\textwidth,clip=}&
\hspace{-6mm}\epsfig{file=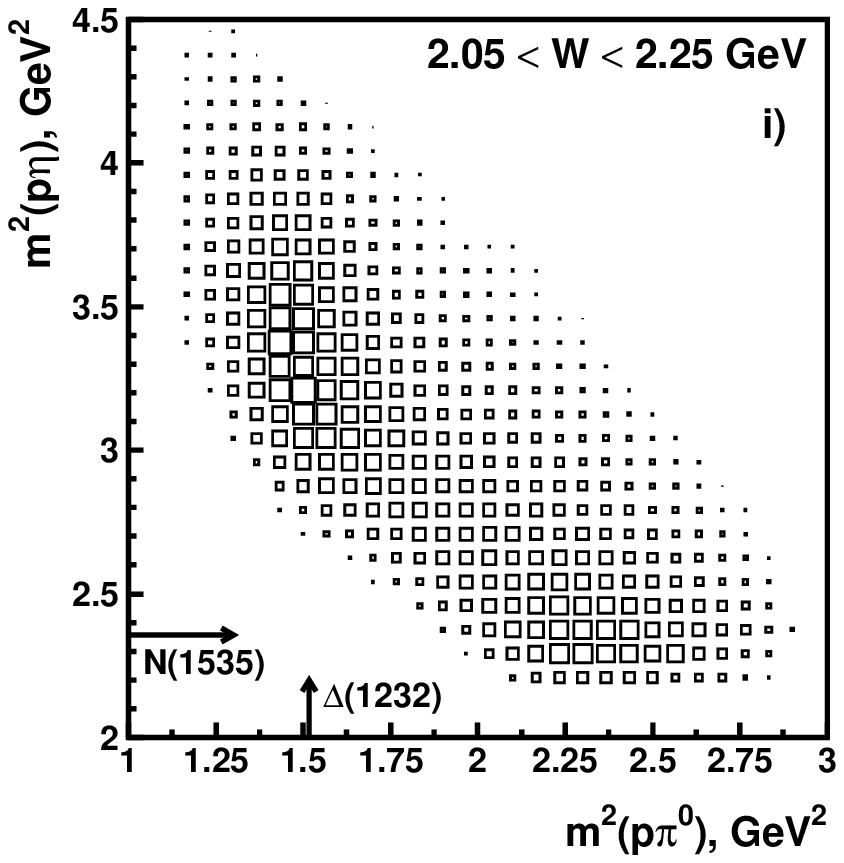,width=0.30\textwidth,clip=}\vspace{-3mm}\\
\hspace{-2mm}\epsfig{file=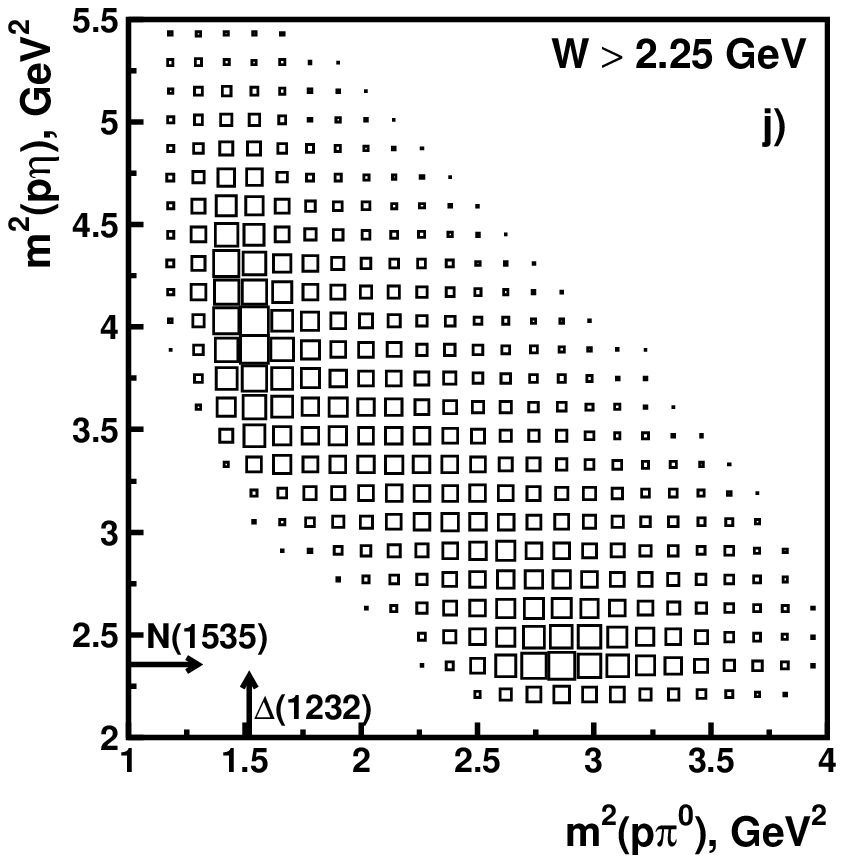,width=0.30\textwidth,clip=}&
\hspace{-6mm}\epsfig{file=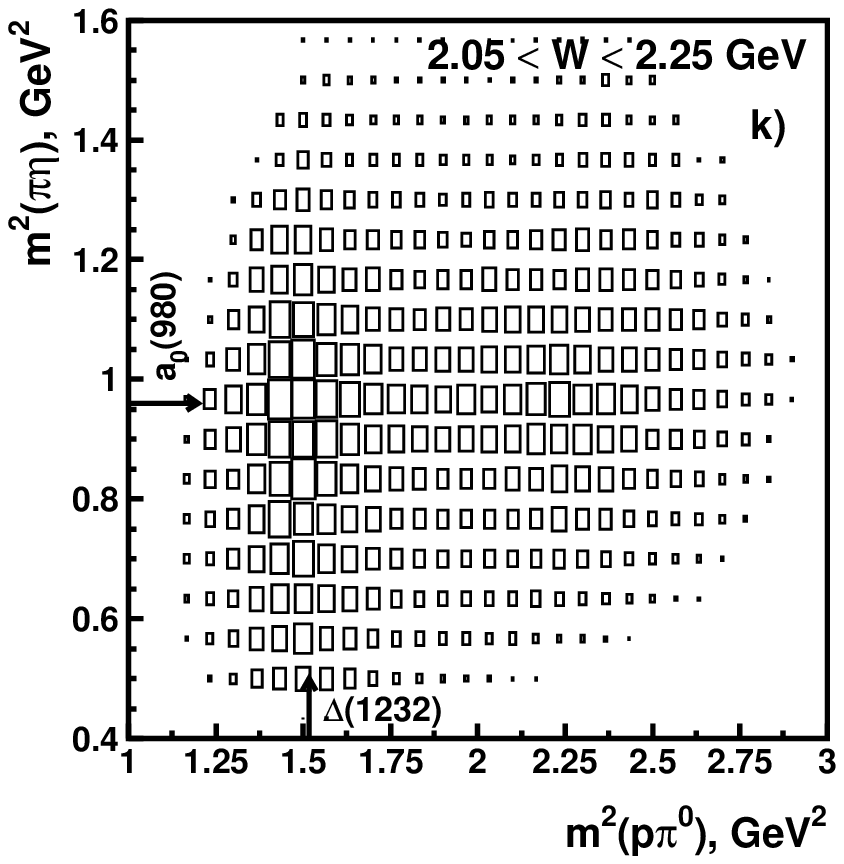,width=0.30\textwidth,clip=}&
\hspace{-6mm}\epsfig{file=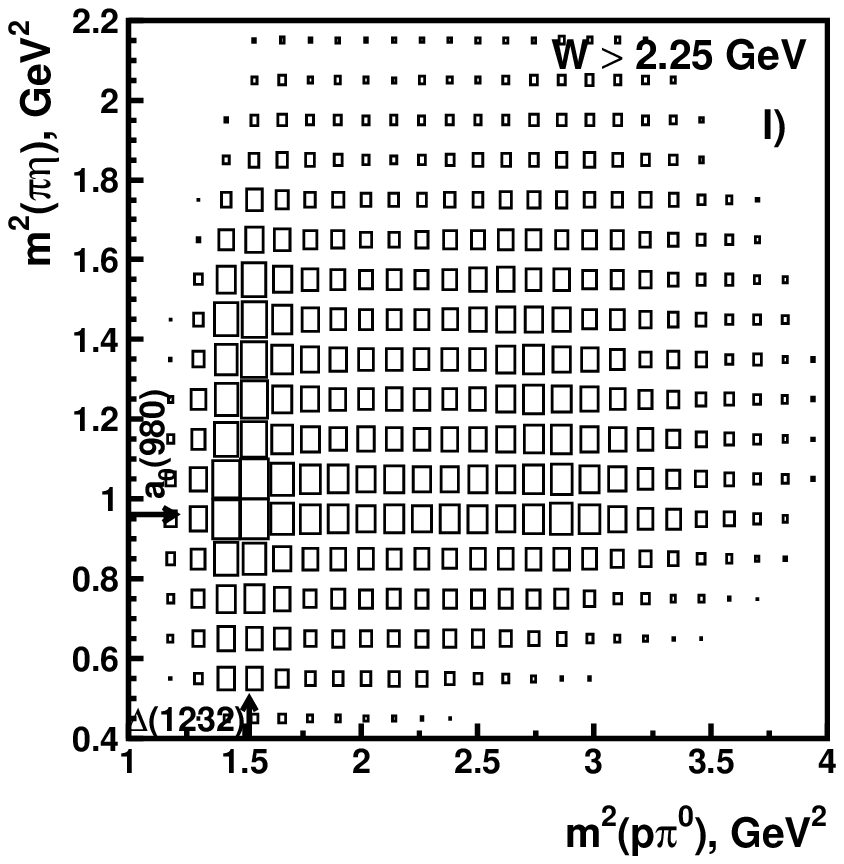,width=0.30\textwidth,clip=}
\end{tabular}
\end{center}\vspace{-2mm}\caption{\label{Dalitz}Dalitz plot for the
reaction $\gamma p\to p\,\pi^0\eta$ for various ranges of the total
energy, data (a-f). The plots (a-d) show $M^2(p\eta)$ versus
$M^2(p\pi^0)$, (e,f) $M^2(\pi^0\eta)$ versus $M^2(p\pi^0)$. With
increasing energy, $\Delta(1232)\eta$ and $N(1535)\pi$ production
fill different kinematical regions and are well separated.
$N(1535)\pi$ is visible only for high photon energies even though
the $N(1535)\pi$ production threshold ($\sim1.0$\,GeV) is lower than
the $\Delta(1232)\eta$ production threshold ($\sim1.2$\,GeV). The
lower part (g-f) show results of the event-based likelihood fit.}
\end{figure*}

\begin{figure*}[pt]
\centerline{
\epsfig{file=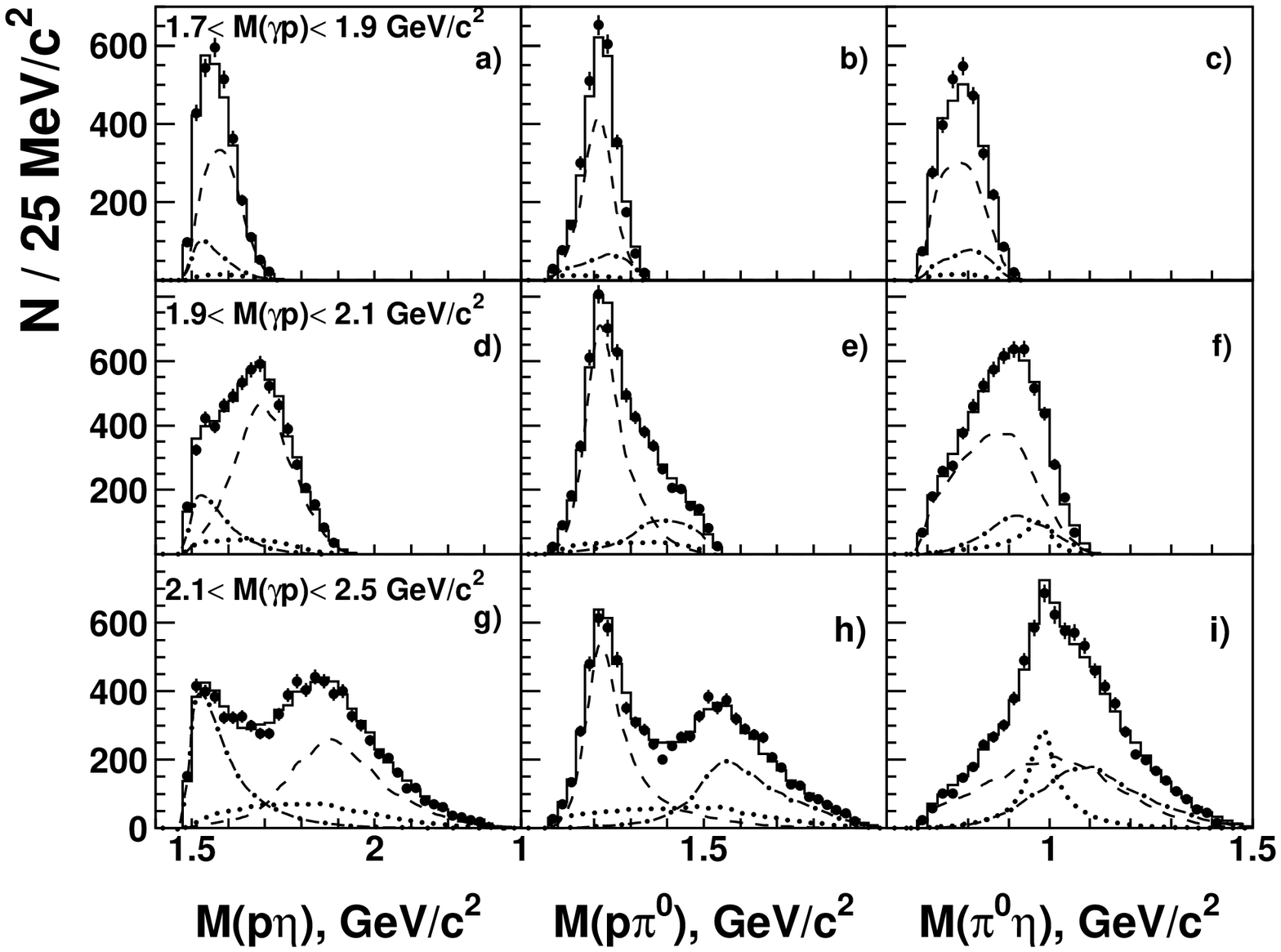,width=0.70\textwidth,clip=}}
\caption{\label{dcs_tot}Mass distributions for three values of
$W=\rm M(\gamma p)$. Data are represented by dots. The distributions
are not corrected for the detector acceptance; errors are of
statistical nature only. The solid curves show the result of the
best fit. The dashed line stands for the $\Delta(1232)\eta$, the
dashed-dotted line for the $N(1535)S_{11}\pi$, and the dotted line
for the $pa_0(980)$ partial wave contribution as determined in the
partial wave analysis. \vspace{2mm}}
 \centerline{
\hspace{8mm}\epsfig{file=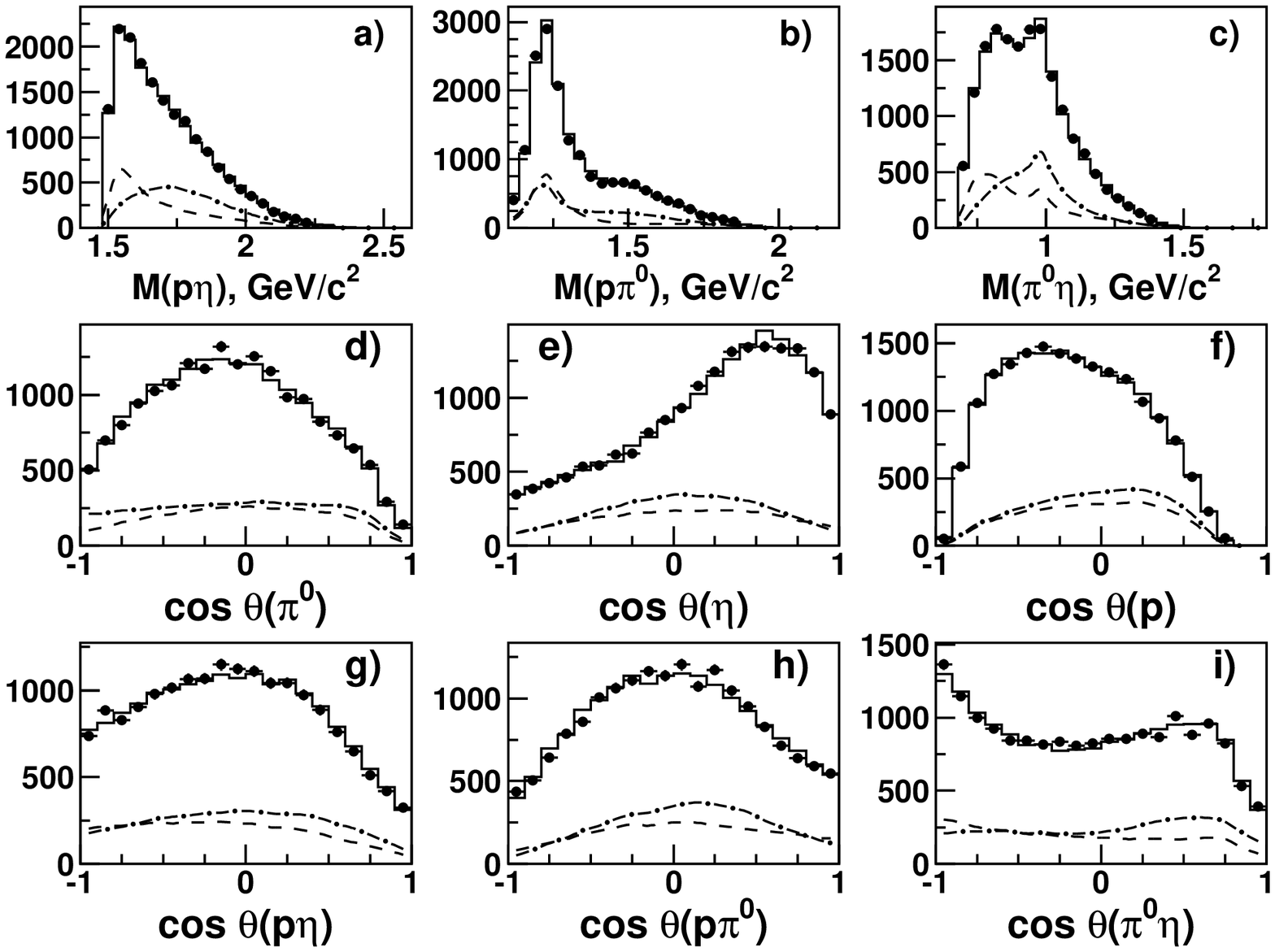,width=0.75\textwidth,clip=}}
\caption{\label{projections}Mass and angular distributions for the
full data set. The distributions are not corrected for acceptance to
allow a fair comparison of the fit with the data without introducing
any model dependence by extrapolating e.g. over acceptance holes.
Data are represented by dots, the fit as solid line. Errors are of
statistical nature only. The dashed line stands for the $D_{33}$,
the dashed-dotted line for the $P_{33}$ partial wave contribution.
There are further PWA contributions which are not shown. (a)
$p\eta$, (b) $p\pi^0$, (c) $\pi^0\eta$ mass distributions. In d)--i)
$\cos\theta$ distributions are shown. In (d-f), $\theta$ is the
angle of a $\pi^0$ (d), $\eta$ (e), $p$ (f) with respect to the
incoming photon in the center--of--mass--system (cms); in (g) the
angle between the $\eta$ and $\pi^0$ in the $p\eta$ rest frame, in
(h) the angle between $\pi^0$ and $\eta$ in the $p\pi^0$ rest frame,
and in (i), the angle between $\pi^0$ and $p$ in the $\pi^0\eta$
rest frame. }
\end{figure*}

\begin{figure*}[pt]
\begin{tabular}{cc}
\epsfig{file=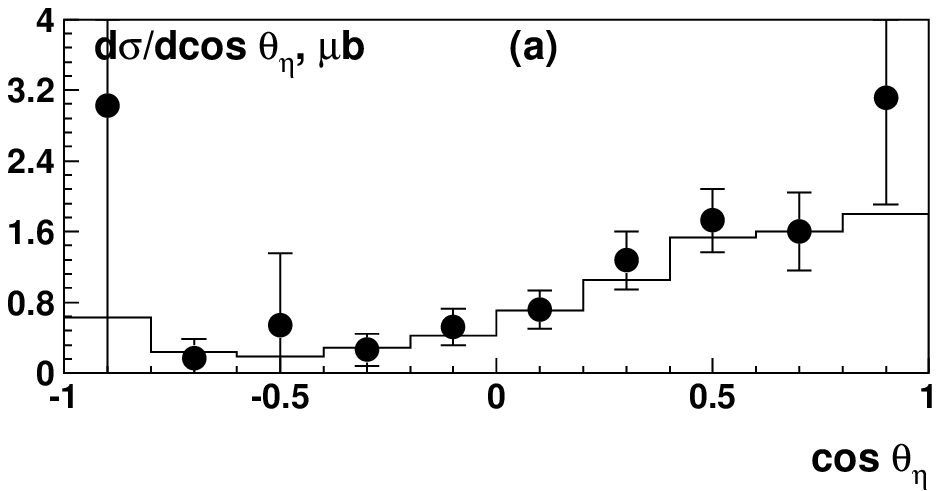,width=0.48\textwidth,clip=}&
\epsfig{file=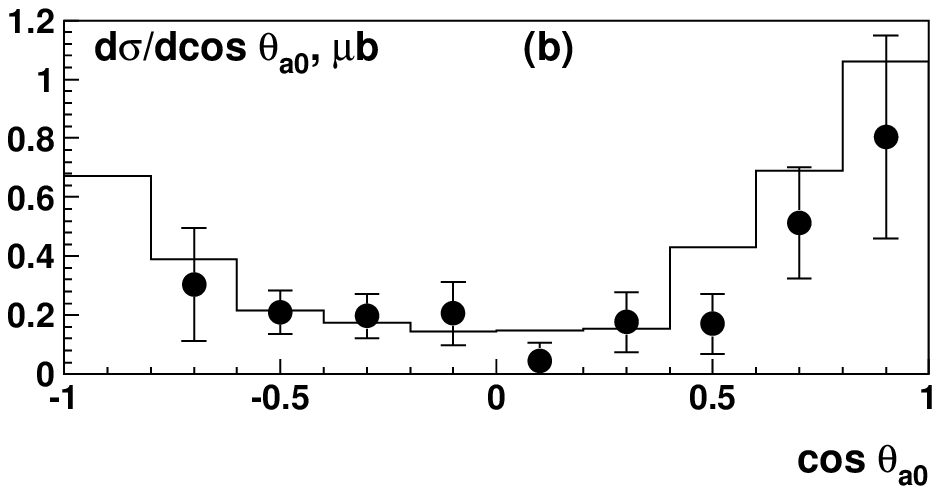,width=0.48\textwidth,clip=}\vspace{-2mm}\\
\epsfig{file=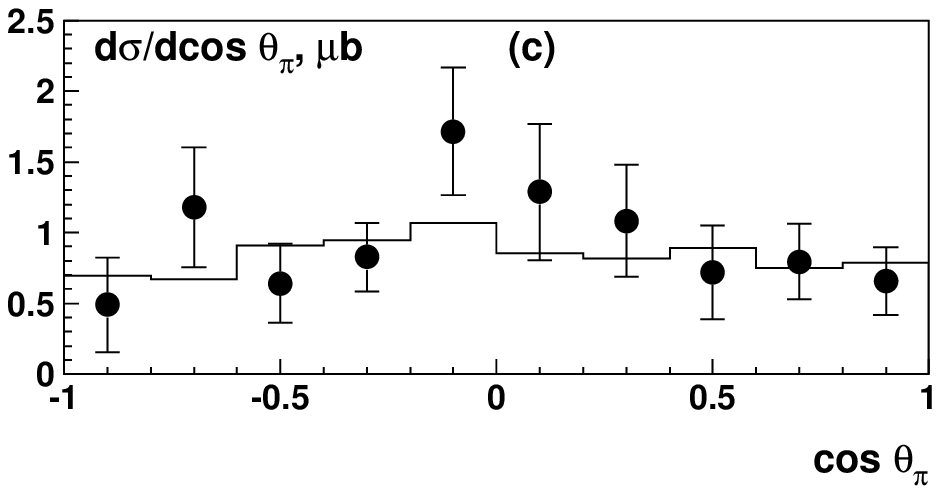,width=0.48\textwidth,clip=}&
\epsfig{file=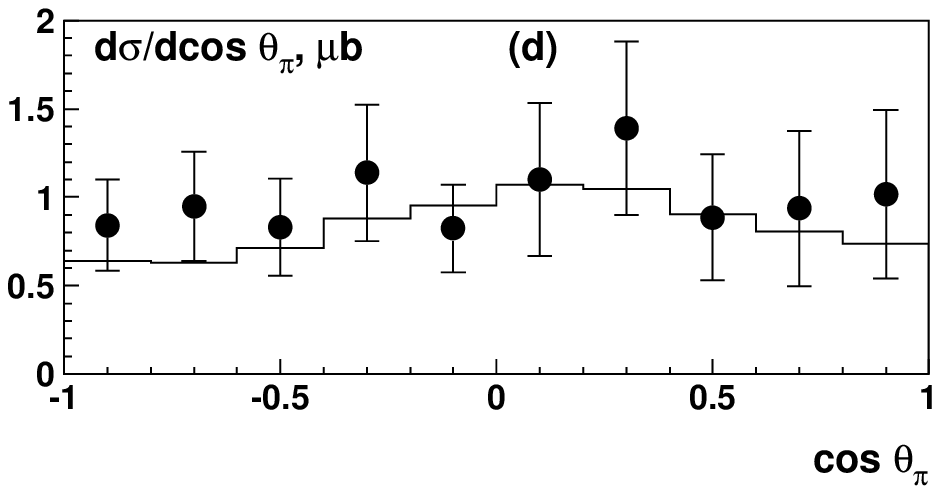,width=0.48\textwidth,clip=}
\end{tabular}
\caption{\label{pion} a: Angular distribution of $\eta$ with respect
to the photon direction in the $\gamma p$ center-of-mass system. b:
Center-of-mass angular distribution of $a_0(980)$ recoiling against
the proton with respect to the photon beam axis for
$E_\gamma>2.0$\,GeV.  c and d: Angular distribution of pions
recoiling in the helicity (c) and Godfrey Jackson (d) frame (after
side bin subtraction).  The latter two distributions are presented
for $1.9 < E_\gamma < 2$\,GeV and $1.15 < M_{p\pi} < 1.30$ to select
$\Delta\eta$ decays, the cuts  $1.075 < M_{p\pi} < 1.150$ and $1.300
< M_{p\pi} < 1.375$\,GeV/c$^2$ define the side bins. Losses due to
the cuts are corrected using Monte Carlo simulations (see text). The
energy region contains large contributions from $\Delta(1920)P_{33}$
and $\Delta(1940)D_{33}$. In all four sub-figures, dots with error
bars give data, the histogram represents the PWA fit.}
\end{figure*}

Fig. \ref{Dalitz} shows a series of $p\,\pi^0\eta$ Dalitz plots for
$M^2(p\eta)$ and $M^2(\pi^0\eta)$ versus $M^2(p\pi^0)$ restricted to
slices in the $\gamma p$ invariant mass $W$, with $W$ covering the
threshold region (a) $1.7<W<1.9$\,GeV/c$^2$, (b) for
$1.9<W<2.05$\,GeV/c$^2$, (c) $2.05<W<2.25$\,GeV/c$^2$, and (d)
$2.25<W<2.5$\,GeV/c$^2$. Figs. (e) and (f) show Dalitz plots for
$M^2(\pi\eta)$ versus $M^2(p\pi^0)$ for the highest energy bin. The
figures in the lower part, (g) to (l), show the same Dalitz plots as
above but calculated from the partial wave fit. All significant
structures in the Dalitz plots are reasonably well reproduced. At
low energies, for $W<2.05$\,GeV/c$^2$, the Dalitz plots are
dominated by $\Delta(1232)\eta$; at high energies, the $N(1535)\pi$
intermediate state is observed. Under these kinematical conditions,
the two contributions populate different kinematical regions and can
be separated easily. The early onset of $\Delta(1232)\eta$ is, at
the first glance, surprising since the threshold for
$\Delta(1232)\eta$ is higher than for $N(1535)\pi$ production.
Within the partial wave analysis this is explained by contributions
from $\Delta(1700)D_{33}$ and $\Delta(1600)P_{33}$. The two
resonances can be produced by an electric or magnetic dipole
transition (and higher-order transitions), respectively, and decay
in $\Delta(1232)\eta$ via S- or P-wave while decays into
$N(1535)\pi$ require P- and D-waves.

Three different 2-body invariant masses can be formed in reaction
(\ref{R1}), the $p\eta$, $p\pi^0$, and $\pi^0\eta$ mass
distributions. They are shown in Fig. \ref{dcs_tot} for three ranges
of photon energies. The distributions undergo significant changes
when the photon energy is varied. In the threshold region, ($1.7\leq
W\leq 1.9$\,GeV/c$^2$), the $p\pi^0$ invariant mass distribution
(Fig. \ref{dcs_tot}b) suggests a significant $\Delta(1232)$
contribution (a presumption confirmed in the partial wave analysis),
the $p\eta$ (Fig. \ref{dcs_tot}a) and $\pi^0\eta$ (Fig.
\ref{dcs_tot}c) mass distributions show no significant structures.
In the 1.9 to 2.1\,GeV/c$^2$ mass region, $\Delta(1232)\eta$ is
still significant, a small $p\eta$ threshold effect is observed
(Fig. \ref{dcs_tot}d). Above 2.1\,GeV/c$^2$, $N(1535)\pi$ -- with
$N(1535)$ decaying into $N\eta$ --  becomes visible (Fig.
\ref{dcs_tot}g). At these energies, $N(1535)$ production is
significant and comparable in intensity to the $\Delta(1232)$. In
the $\pi^0\eta$ mass distribution, a peak is observed at
1\,GeV/c$^2$ (Fig. \ref{dcs_tot}i) which we assign to the
$pa_0(980)$ intermediate state. Mass projections for the full data
set and some angular distributions are presented in Fig.
\ref{projections}. The $P_{33}$ and $D_{33}$ partial wave
contribution are also shown.

We now turn to the discussion of a few specific angular
distributions. In the threshold region for $p\pi^0\eta$
photoproduction we may expect $S$-and $P$-waves to dominate the
decay mode. $S$-wave decays into $\Delta(1232)\eta$ require $D_{33}$
quantum numbers which can be reached by an electric dipole
transitions; magnetic dipole transitions lead to $P_{33}$ quantum
numbers for which $P$-wave decays into $\Delta(1232)\eta$ are
possible. If one of the two amplitudes prevails, we may expect an
isotropic angular distribution for the recoiling pion. In case of
both amplitudes being present, $S$-$P$ interference leads to a
linear increase with $\cos\theta_{\eta}$, a distribution which is
clearly contributing to the data (Fig. \ref{pion}a).

At the largest photon energies, a peak due to $a_0(980)$ production
can be identified in the $\pi\eta$ invariant mass distribution. In
Fig. \ref{pion}b the $a_0(980)$ angular distribution with respect to
the photon beam is presented (after side bin subtraction). The
$a_0(980)$ signal region is defined by $0.96 < M_{\pi\eta} <
1.04$\,GeV/c$^2$, the sidebins with $0.88 < M_{\pi\eta} < 0.96$ and
$1.04 < M_{\pi\eta} < 1.12$ are subtracted with weight 1/2. The side
bin subtraction suffers from the large $a_0(980)$ width, hence the
distribution must be viewed with some precaution. However, it agrees
reasonably well with the PWA result and the subtraction is justified
a posteriori.

Surprisingly, $a_0(980)$ is not mostly produced in forward direction
as might be expected if the main $a_0(980)$ production mechanism
would have been vector-meson exchange in the $t$-channel (exploiting
a $a_0(980)\to \gamma\,\rho /\omega$ coupling). A significant
fraction of the $a_0(980)$ signal seems to stem from the decay of
high-mass resonances. This suspicion is supported by the partial
wave analysis.

Fig. \ref{pion}c and d show $\Delta(1232)\to p\pi^0$ decay angular
distributions which should reflect the spin $J=3/2^+$ quantum
numbers of $\Delta(1232)$. The $\pi^0$ angular distribution from
$\Delta$ decays is shown as pion emission angle with respect to the
$\Delta$ flight direction (helicity frame, Fig. \ref{pion}c), or
with respect to the photon beam direction (Godfrey-Jackson frame,
Fig. \ref{pion}d). The photon energy is restricted to
$W<1.9$\,GeV/c$^2$. The distributions are essentially flat
(preventing a straightforward identification of the $\Delta(1232)$
spin) indicating that $\Delta(1232)$ are likely not produced by one
helicity amplitude only and in a single reaction chain. Explicit
expressions for typical angular distributions are given elsewhere
\cite{Anisovich:2006bc}.

\section{Partial wave analysis}
\subsection{The method}
The qualitative findings discussed in the previous section are
confirmed in a partial wave analysis (PWA). The formalism is
documented in
\cite{Anisovich:2004zz,Anisovich:2006bc,Anisovich:2007ra}. In
addition to the data presented here, data on photoproduction of
single pions, $\eta$'s, and of $\Lambda$ and $\Sigma$ hyperons are
included as well as some important partial-wave amplitudes for $\pi
N$ elastic scattering. References to the data, an outline of the PWA
method and the definition of total likelihood and likelihood
contributions can be found in \cite{Anisovich:2007bq}. The different
data sets entered with appropriately chosen weights $w_i$; the
weights vary from 1 (for data which should be reproduced
qualitatively) to 30 (for low-statistics data like polarization data
which we insist to be well reproduced in the fits). The weights are
given in Table 4 of \cite{Anisovich:2007bq}. The weights are chosen
to ensure that there is good qualitative agreement between data and
fit even for low-statistics data which otherwise may be dominated by
the influence of high-statistics data. The mean weight is $\overline
w=4.2$. For this data, $w_{\gamma p\to p\,\pi^0\eta}=10$ was chosen;
(much) smaller weights lead to discrepancies between fit and this
data, higher weights deteriorate the fits elsewhere. We minimized a
pseudo-log-likelihood function $\ln\mathcal L_{\rm tot}$ defined as
a sum of the log-likelihoods of fits to $\gamma p\to \pi^0\pi^0$ and
$\gamma p\to \pi^0\eta$, and of the $\chi^2/2$ contributions of all
data given in the form of histograms.

The $\gamma p\to p\,\pi^0\pi^0$ and $\gamma p\to p\,\pi^0\eta$ data
are fitted using an event-based maximum likelihood method which
takes into account the full correlations between all variables in
the 5-dimensional phase space. New data on the beam asymmetry
$\Sigma$ for $\gamma p\to p\,\pi^0\eta$ \cite{Gutz:2008} were
included in the partial wave analysis. Since the prediction for
$\Sigma$ were already close to the data \cite{Gutz:2007}, the new
data on $\Sigma$ did not change the PWA results.

\begin{table}[pt]
\caption{\label{isobars}The lowest partial waves which may
contribute to $\gamma p\to p\,\pi\eta$.\vspace{2mm}}
\begin{center}
\renewcommand{\arraystretch}{1.4}
\hspace{-3mm}\begin{tabular}{cccc} \hline\hline L
\hspace{-2mm}&\hspace{-3mm}
$\Delta(1232)\eta$            \hspace{-3mm}&\hspace{-3mm} $N(1535)\pi$ \hspace{-3mm}&\hspace{-3mm} $pa_0(980)$\\
\hline 0\hspace{-3mm}&\hspace{-3mm}$D_{33}$\hspace{-3mm}&\hspace{-3mm}$P_{11}$, $P_{31}$\hspace{-3mm}&\hspace{-3mm}$P_{11}$, $P_{31}$ \\
1\hspace{-3mm}&\hspace{-3mm}$P_{31}$, $P_{33}$,
$F_{35}$\hspace{-3mm}&\hspace{-3mm}$S_{11}$, $D_{13}$, $S_{31}$,
$D_{33}$\hspace{-3mm}&\hspace{-3mm}$S_{11}$, $D_{13}$, $S_{31}$, $D_{33}$           \\
2\hspace{-3mm}&\hspace{-3mm}$S_{31}$, $D_{33}$, $D_{35}$,
$G_{37}$\hspace{-3mm}&\hspace{-3mm}$P_{13}$, $F_{15}$,
$P_{33}$, $F_{35}$\hspace{-3mm}&\hspace{-3mm}$P_{13}$, $F_{15}$, $P_{33}$, $F_{35}$ \\
 \hline\hline
\end{tabular}
\renewcommand{\arraystretch}{1.0}
\end{center}
\end{table}

As a starting point, we used the set of resonances of
\cite{Anisovich:2007bq} and added resonances in the partial waves
listed in Table \ref{isobars}. The Table gives the quantum numbers
of baryon resonances which could contribute to the $p\pi\eta$ final
state assuming that the relative orbital angular momenta between
meson and baryon in the three visible intermediate states
$\Delta(1232)\eta$, $N(1535)\pi$, and $pa_0(980)$ are restricted to
$L\leq 2$. First, resonances of the first line were tested but their
decays with $L=1$ and $L=2$ were also admitted. It turned out that
for nearly all resonances, inclusion of $L=2$ decays led to marginal
likelihood improvements ($\Delta\ln\mathcal L_{\pi^0\eta}<5$
corresponding to a statistical significance of less than two
standard deviations. The $\Delta(1232)\eta$ D-wave contribution,
e.g., was found to be negligible for $\Delta(1700)D_{33}$, below 1\%
for $\Delta(1940)D_{33}$ while for $\Delta(2360)D_{33}$, the $S$ and
$D$ wave couplings had similar magnitudes. The partial waves in the
second (and third) line of Table \ref{isobars} were tested one by
one. No significant contribution was found except for a small
fraction of $\Delta(1905)F_{35}$ in its $\Delta(1232)\eta$ decay.
Its fraction in $N(1535)\pi$ (with $L=2$) is marginal.

The quality of the final PWA fit to the data is shown in Fig.
\ref{dcs_tot} and \ref{projections}. In the latter figure, the full
statistics integrated over the full energy range is shown; the
agreement between data and fit result is very convincing and equally
good when mass slices are selected or other variables are plotted.
It has to be remembered that a likelihood fit uses the full
kinematics of each event and that neither the  Dalitz plots
(Fig.~\ref{Dalitz}) nor the projections (Fig.~\ref{dcs_tot}) are
fitted.

\subsection{Isospin considerations}

\begin{figure*}[pt]
\vspace{4mm}
 \centerline{
\epsfig{file=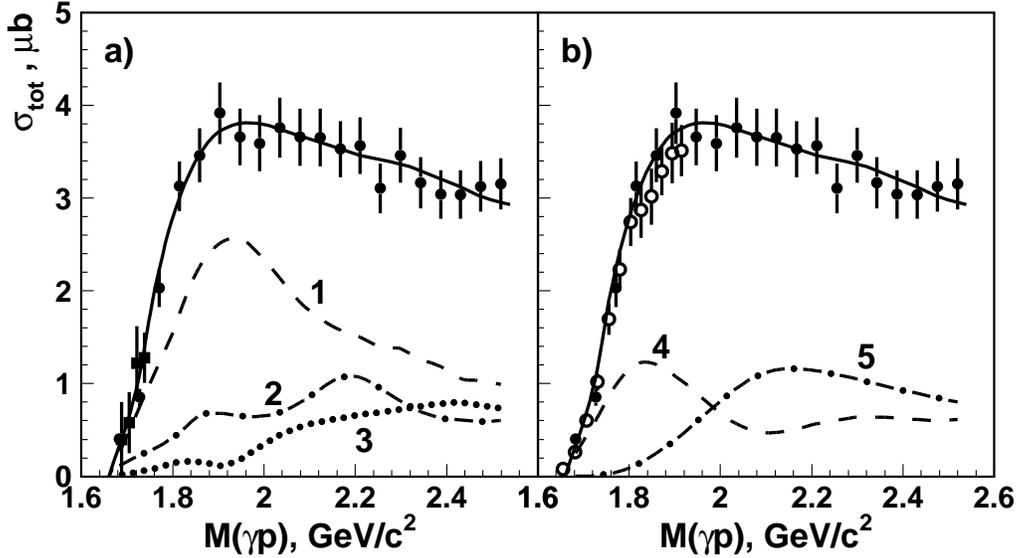,width=0.75\textwidth,clip=}}
\caption{\label{tot_pwa}Total cross sections for $\gamma p \to
p\,\pi^0\eta$. In both panels, our data are represented by $\bullet$
with error bars; the solid curves represents our PWA fit. Comparison
to other data and two decompositions of the total cross sections are
shown in two sub-figures. a) Our data compared to data from
Nakabayashi {\it et al.} \cite{Nakabayashi:2006ut} {\tiny
$\blacksquare$}. Curve 1 shows the cross section for $\gamma p\to
\Delta(1232)\,\eta$, curve 2 that for $N(1535)\pi$, and curve 3 that
for $\gamma p\to p\,a_{0}(980)$. These cross sections are not
corrected for unseen decay modes. b) open circles: Ajaka  {\it et
al.} \cite{Ajaka:2008}. Partial wave decomposition of the cross
section into the $D_{33}$ (curve 4) and the $P_{33}$ partial wave
(curve 5). }
\end{figure*}
Some resonances, even though known to exist, cannot be reasonably
included in the partial wave analysis, for technical reasons. We
need e.g. a $J^P=1/2^+$ resonance at about 1900\,MeV/c$^2$. A
$N(1880)P_{11}$ resonance (formerly called $N(1840)$ by us
\cite{Sarantsev:2005tg}) is required to describe the two reactions
$\gamma p\to \Lambda\,K^+$ and $\gamma p\to \Sigma\,K^+$ (which
define the isospin of contributing resonances). Both,
$N(1880)P_{11}$ and $\Delta(1910)P_{31}$, can decay into
$N(1535)\pi$ and/or $pa_0(980)$, and there is no information in
these two decay modes to decide on the isospin. Hence both might be
present. The $\Delta(1910)P_{31}$ might decay into
$\Delta(1232)\eta$ but this is not observed. Thus there is evidence
for $N(1880)P_{11}$ and no direct evidence for an additional
$\Delta(1910)P_{31}$. However, a $\Delta(1910)P_{31}$ might
nevertheless exist and contribute to $p\pi^0\eta$ if it couples to
$N(1535)\pi$ and/or $pa_0(980)$ but not to $\Sigma\,K^+$ and
$\Delta(1232)\eta$ (or only with a weak coupling). In such a case
(which admittedly is a bit constructed but not forbidden by any
selection rule), $\Delta(1910)P_{31}$ decays into $p\pi^0\eta$ are
discarded. The reason for this is that introduction of two close-by
resonances in one final state having different isospin but otherwise
identical quantum numbers leads very often to very large amplitudes
for both resonances and destructive interference.

The $N(2200)P_{13}$ was introduced, too, in \cite{Sarantsev:2005tg}.
Here, it is found to contribute significantly to $p\,\pi^0\eta$ but
not to $\Delta(1232)\eta$. Hence we assume it to belong to the
familiy of nucleon resonances, and do no include a further
$\Delta(2200)P_{33}$. For the same reason we do not allow
$N(1875)D_{13}$ to decay into $N(1535)\pi$ and/or $pa_0(980)$; in
the 3/2$^-$ partial wave, the intensity in $p\pi^0\eta$ is assigned
to $\Delta(1940)D_{33}$. The isospin $I=3/2$ is identified from the
$\Delta\eta$ decay mode. Contributions from $I=1/2$ to $N(1535)\pi$
and/or $pa_0(980)$ are possible but are not considered here because
of limitations of the data base. A separation of two isospin
contributions would require a second isospin channel, e.g., the
study of $\gamma p\to n\pi^+\eta$.

\subsection{The total cross section}

The final fit required contributions to the $p\to p\pi^0\eta$ final
state from eight resonances; these are listed in Table
\ref{pwa-list}.
\begin{table}[pb]
\vspace{-2mm}\caption{\label{pwa-list}Resonances used in the partial
wave analysis. \vspace{2mm}}
\begin{center}
\renewcommand{\arraystretch}{1.4}
\begin{tabular}{cccc}
\hline\hline $N(1880)P_{11}$& $N(2200)P_{13}$& $\Delta(1600)P_{33}$
&$\Delta(1920)P_{33}$  \\
$\Delta(1700)D_{33}$& $\Delta(1940)D_{33}$& $\Delta(2360)D_{33}$ &
$\Delta(1905)F_{35}$\\
 \hline\hline
\end{tabular}
\renewcommand{\arraystretch}{1.0}
\end{center}
\vspace{-2mm}\end{table} Two of them do not couple to
$\Delta(1232)\eta$ and are interpreted as nucleon resonances. The
$N(1880)$ could be the missing $P_{11}$ resonance forming, jointly
with $N(1900)P_{13}$, $N(2000)F_{15}$, $N(1990)F_{17}$, a
super-multiplet with (dominantly) $L$=2, $S$=3/2. The
$N(2200)P_{13}$ might be a $N(1900)P_{13}$ radial excitation.

In Fig.~\ref{tot_pwa} the total cross section is displayed. The
points with errors give the acceptance-corrected results of the
measurement and their statistical and systematic errors but not the
$\pm15$\% systematic error assigned to the photon-flux
normalization. The solid curve shows the result of the partial wave
analysis (PWA). The total cross section reaches the maximum of
4\,$\mu$b in the 2\,GeV/c$^2$ region and then decreases slowly to
3\,$\mu$b. In the threshold region, our cross section is fully
compatible with results obtained at Sendai
\cite{Nakabayashi:2006ut}. The cross section determined by GRAAL
\cite{Ajaka:2008} falls systematically below our values but the
difference is covered by the 15\% normalization error.

The acceptance is calculated using Monte Carlo techniques. The full
apparatus is simulated using GEANT3. Detection and reconstruction
efficiency, probabilities for split-offs for photons and charged
particles, confidence level distributions for the different
kinematical fits for real data and Monte Carlo data were compared
carefully. The detector is well understood by the Monte Carlo
simulation \cite{van Pee:2007tw}. The mean acceptance for the
$p4\gamma$ final state is 15-20\% but the acceptance vanishes for
forward protons escaping through the forward hole in the detector.
Hence a proper extrapolation of the four-dimensional event
distribution into the uncovered region of the detector is essential.

For this purpose, phase space distributed events due reaction (1)
were simulated according to the photon flux. These events were
weighted with the PWA solution. The resulting event sample
represents the `true' physics, in the limit that data and partial
wave analysis are correct. The curves in Figures \ref{pion} and
\ref{tot_pwa} are calculated as Monte Carlo integral over this
(weighted) event sample. The detection efficiency for the data is
obtained from the ratio of reconstructed and generated events of the
weighted Monte Carlo event sample.

The systematic error comprises the error in the reconstruction
efficiency (5.7\%) and a few small errors ($\approx 1$\%) due to
uncertainties in target position and profile of the photon beam, and
the extrapolation of the intensity distribution into blind regions
of the detector (mainly for forward protons). The latter uncertainty
is estimated to about 5\% by using different PWA solutions giving
acceptable descriptions of the data. These errors are added in
quadrature and included in Fig. \ref{tot_pwa}. The largest error
stems from the uncertainty in the photon flux: the normalization
derived from a comparison of the $\pi^0$ photoproduction cross
section with SAID (see \cite{van Pee:2007tw}). The normalization
error of $\pm 15$\% is not included.

The excitation functions for the three intermediate states
$\Delta(1232)\eta$, $N(1535)\pi$, and $pa_0(980)$, deduced from the
partial wave analysis, are shown in Fig. \ref{tot_pwa}a. The
$\Delta(1232)\eta$ makes the most significant contribution to the
total cross section; at threshold, it dominates the reaction. It
reaches a peak height just below 2\,GeV/c$^2$. The $N(1535)\pi$
intermediate state exhibits a structure at 1.9\,GeV/c$^2$ (due to
the $N(1880)P_{11}$) and a second bump at 2.2\,GeV/c$^2$ which we
interpret as $N(2200)P_{13}$. With increasing mass, $pa_0(980)$
gains a notable intensity. The excitation functions shown in Fig.
\ref{tot_pwa} represent the best solution. It should be stressed
that these contributions are qualitative; the numerical values
differ significantly for different fits, much more than the pole
positions of the contribution resonances. The errors in the
excitation functions can be, in particular at high energies, as
large as 30\%.

\subsection{Resonances contributing to \boldmath$p\pi^0\eta$}

In a first approach, we used Breit-Wigner resonances for less
important waves while $S_{11}$, $P_{11}$, $P_{13}$, $P_{33}$, and
$D_{33}$ were described within the K-matrix/P-vector approach
\cite{Anisovich:2007bq}. For these waves, elastic $\pi N$ scattering
amplitudes from \cite{Arndt:2006bf} were included in these fits. For
the two waves $P_{33}$ and $D_{33}$ discussed here, a satisfactory
description was obtained with a 6-channel ($N \pi$,
$\Delta(1232)\pi$ ($S,D$-waves for $D_{33}$ and $P,F$-waves for
$P_{33}$), $\Delta(1232)\eta$, $N(1535)\pi$, $N\,a_0(980)$) 3-pole
parameterization. Parameterizations of the $S_{11}$,  $P_{11}$, and
$P_{13}$ amplitudes are given in \cite{Anisovich:2007ra}, those for
$P_{33}$ and $D_{33}$ are presented in Tables \ref{Table:p33_km} and
\ref{Table:d33_km}.

\begin{table}[pt]
\caption{\label{Table:p33_km} Properties of the $P_{33}$ $K$-matrix
poles. The masses $M_i$ and $K$-matrix coupling constants
$g^{(i)}_{\alpha}$ from resonance $(i)$ to the decay mode $\alpha$
are given in GeV, the helicities $A$ in GeV$^{-\frac{1}{2}}$, $s$ in
GeV$^2$/c$^4$; the $f$ - describing $N\pi\to\Delta\pi$ contact
interactions - are dimensionless. The parameters marked with $^*$
were fixed in the fit.}
\begin{center}
\renewcommand{\arraystretch}{1.4}
\begin{tabular}{clcccc}
\hline\hline
 &                     &Pole 1       &Pole 2       &Pole 3       &        \\
\hline
 &$M_i$                  &  1232$\pm$3&  1795$\pm$40&  2020$\pm$80&
 \\
 &$A_{1/2}$            & -0.125      & 0.031      &  0.026      &
 \\
 &$A_{3/2}$            & -0.264      & 0.123      &  0.081      &
 \\
\hline
   $ a $&              &$g^{(1)}_a$  &$g^{(2)}_a$  &$g^{(3)}_a$  &$f_{1a}$\\
\hline
 1&$N(940)\pi $          &  1.20    &  1.25  &  0.92  & -1.40   \\
 2&$\Delta(1232)\pi(P)$  & -0.52    & -1.73  & -0.69  &  0.81   \\
 3&$\Delta(1232)\pi(F)$  &  0.$^*$  &  0.19  & -0.76  &  0.$^*$   \\
 4&$\Delta(1232)\eta  $  &  0.$^*$  &  0.75  & -0.63  &  0.$^*$   \\
 5&$N(1535)\pi       $   &  0.$^*$  &  0.94  &  1.04  &  0.$^*$   \\
 6&$N(940)\rho        $  &  0.$^*$  &  0.05  & -0.06  &  0.$^*$   \\
\hline\hline
\end{tabular}
\renewcommand{\arraystretch}{1.0}
\end{center}
\caption{\label{Table:d33_km} Properties of the $D_{33}$ $K$-matrix
poles. \vspace{-2mm}}
\begin{center}
\renewcommand{\arraystretch}{1.4}
\begin{tabular}{clcccc}
\hline\hline
 &                     &Pole 1       &Pole 2       &Pole 3       &        \\
\hline
 &$M_i$                  &  1780$\pm$50&  1980$\pm$60&  2400$\pm$90&
 \\
 &$A_{1/2}$            & +0.041      & 0.205      &  0.228      &
 \\
 &$A_{3/2}$            &  0.252      & -0.003      &  0.143      &
 \\
\hline
   $ a $&              &$g^{(1)}_a$  &$g^{(2)}_a$  &$g^{(3)}_a$  &$f_{1a}$\\
\hline
 1&$N(940)\pi $          &  0.31  &  0.66  &  1.22  & -0.62   \\
 2&$\Delta(1232)\pi(S)$  &  0.51  & -0.72  & -0.16  & -0.17   \\
 3&$\Delta(1232)\pi(D)$  & -1.21  & -1.49  & -1.51  & -0.63   \\
 4&$\Delta(1232)\eta  $  & -0.39  & -0.23  &  0.81  &  0.$^*$   \\
 5&$N(1535)\pi        $  & -0.38  & -0.40  & -0.78  &  0.$^*$   \\
 6&$N(940)\rho        $  & -0.14  & -0.26  & -0.24  &  0.$^*$   \\
\hline\hline
\end{tabular}
\renewcommand{\arraystretch}{1.0}
\vspace{-1mm}
\end{center}
\end{table}
\begin{figure*}[pt]
\epsfig{file=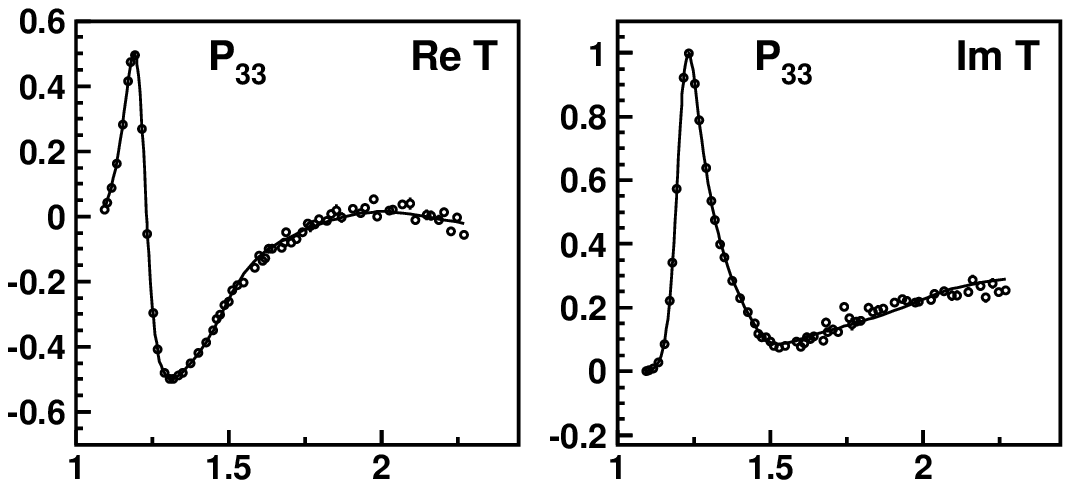,width=0.5\textwidth,clip=}
\epsfig{file=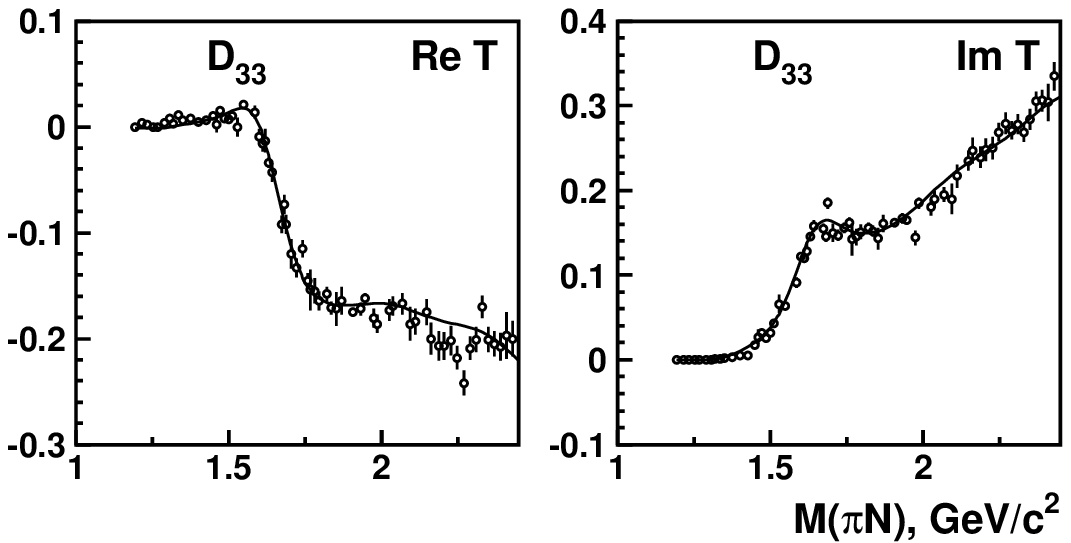,width=0.5\textwidth,clip=}
\caption{\label{pwad33} The $\pi N$ elastic scattering amplitude in
the $P_{33}$ (left) and $D_{33}$ (right) wave \cite{Arndt:2006bf}
and our fit. The $P_{33}$ shows the distinct phase motion for
$\Delta(1232)$; a second resonance at 1600\,MeV/c$^2$ is difficult
to see, and there is no sign for a third resonance. Likewise, there
is a rapid phase motion at 1700\,MeV/c$^2$ in the $D_{33}$ but no
sign for any resonance above it.}
\end{figure*}

\begin{table*}[pt]
\caption{\label{pwa}Resonances contributing to $\gamma p\to
p\,\pi\eta$, their pole positions and decay modes, and change of
likelihood when resonance couplings to $N\pi\eta$ are reduced to
zero. A change in likelihood by 50 corresponds to a $\chi^2$ change
of 25 or to a statistical significance of $5\,\sigma$. \vspace{2mm}}
\begin{center}
\renewcommand{\arraystretch}{1.4}
\begin{tabular}{cccccccccc}
\hline\hline Name&Wave& Pole (MeV/c$^2$)& Decay modes
&&\hspace{-6mm}
Fraction&$\Delta\ln\mathcal L_{\rm tot}$ & $\Delta\ln\mathcal L_{\pi^0\eta}$\\
\hline
$N(1880)$&$P_{11}$& $~1880-i110$ & $pa_0, N(1535)\pi$&&\hspace{-6mm}~12\% & 2050&70\\
$N(2200)$&$P_{13}$& $~2200-i150$ & $pa_0, N(1535)\pi$&&\hspace{-8mm} ~7\%&1392&73\\
$\Delta(1600)$& $P_{33}$ & $1510-i115$ & $\Delta\eta$
&\hspace{-4mm}\multirow{2}{2mm}{\huge\}}&\hspace{-11mm}\multirow{2}{1mm}{~22\%}&884&60\\
$\Delta(1920)$&$P_{33}$ & $1980-i175$ & $\Delta\eta, pa_0$ && &1818&63 \\
$\Delta(1700)$& $D_{33}$ & $1640-i160$ & $\Delta\eta, N(1535)\pi$
&\hspace{-4mm}\multirow{3}{2mm}{\huge\}} &&2611&84 \\
$\Delta(1940)$&$D_{33}$&$1985-i195$&$\Delta\eta, pa_0, N(1535)$&&\hspace{-6mm}~24\%&1124&52\\
$\Delta(2360)$&$D_{33}$ & $2320-i250$ & $\Delta\eta, pa_0, N(1535)$&&&1018&24 \\
$\Delta(1905)$& $F_{35}$&$1920-i145$ &$\Delta\eta, N(1535)\pi$&&
\hspace{-6mm}~7\%&1991&65\\
\multicolumn{5}{c}{$t$-channel $\rho$-$\omega$-,
$u$-channel $p$-$\Delta$-exchange} &\hspace{-6mm}~28\%&6340&208\\
 \hline\hline
\end{tabular}
\renewcommand{\arraystretch}{1.0}
\end{center}
\vspace{-4mm}
\end{table*}

All eight resonances are required to get a good fit. Table \ref{pwa}
lists the resonances used in the final fits (column 1 and 2), their
pole positions (column 3) and their decay modes leading to the
$p\pi^0\eta$ final states (column 4). The fractional contributions
(adding up to 100\% of the integrated $p\pi^0\eta$ cross section)
are listed in column 5, while columns 6 and 7 give the likelihood
change, for the total data set and for the data presented here,
respectively, when a resonance is assumed not to couple to
$p\pi^0\eta$.

In the $P_{33}$ wave we find a first pole above $\Delta(1232)$ at
$1510^{+20}_{-50}-i\,115\pm 20$\,MeV/c$^2$ and a further pole at
$1980^{+25}_{-45}-i 175^{+18}_{-28}$\,MeV/c$^2$. If the couplings of
the first pole to $p\pi\eta$ are reduced to zero, the likelihood for
the $\gamma p\to p\pi\eta$ data deteriorates by 60. Removing
$p\pi\eta$ couplings of the highest K-matrix pole ``costs'' 63 units
in likelihood. If the latter pole is completely removed from the
fit, the likelihood value for $\pi\eta$ channel changes by 177 and
total likelihood by 2230. Such fits are unacceptable.

Using only $\Delta(1700)D_{33}$ for the $D_{33}$ wave, no
satisfactory description of the data was obtained; a second
resonance proved to be essential. The pole position for the second
resonance was found at $(1985\pm 35)\,-i\,(195\pm 25)$ MeV/c$^2$.
The likelihood value for the $\gamma p\to p\,\pi^0\eta$ reaction
changed by $\Delta \ln \mathcal L_{\gamma p\to p\,\pi^0\eta}$ =151,
the total likelihood by 2005 units; these are highly significant
numbers. With $\Delta(1940)D_{33}$ and without $\Delta(1700)$
$D_{33}$, the fit was also unacceptable ($\Delta\ln\ \mathcal
L_{\gamma p\to p\,\pi^0\eta}$ =137, $\Delta \ln \mathcal L_{\rm
tot}$ =4782). The fit improved notably ($\Delta\ln\ \mathcal
L_{\gamma p\to p\,\pi^0\eta}$ =28, $\Delta \ln \mathcal L_{\rm
tot}$=770) by adding a third $D_{33}$ state. Its mass optimized at
about 2.35\,GeV/c$^2$ and a width in the 400-600\,MeV/c$^2$ range.
Its mass is close to the end of the available phase space. An upper
bound for its mass is not well defined; hence we do not claim that
this state exists.

When an additional $N(2000)D_{13}$ resonance decaying into
$N(1535)\pi$ and $Na_0(980)$ was introduced, strong interference
between the two amplitudes was observed which increases the
systematical errors of the PWA results. \ A precise mass
determination of the $\Delta(1940)D_{33}$ will hence require
separation of the two isotopic resonances, e.g. by measurements of
$\gamma N\to n\pi^+\eta$. This ambiguity affects of course the
branching ratios as well.

We assign the absence of $\Delta(1920)P_{33}$ and
$\Delta(1940)D_{33}$ in the phase shift analysis of Arndt {\it et
al.} \cite{Arndt:2006bf} to their weak couplings to the $N\pi$
channel. To check if our data are consistent with elastic scattering
amplitudes, we included these amplitude in the fit. The result is
shown in Fig. \ref{pwad33}: the existence of the two resonances is
not in conflict with elastic scattering.

Fig. \ref{Dalitz}e shows a structure in the $p\pi^0$ invariant mass
at about 1.5\,GeV (while Fig. \ref{Dalitz}f would suggest a higher
mass). We tested the possibility that the structure signals
contributions from $\gamma p\to N(1520)\eta$ or $\gamma p\to
\Delta(1600)\eta$. Both attempts yielded marginal improvements in
likelihood and contributions of about 0.5\%. The PWA interprets the
structure as interference between $pa_0(980)$ and $N(1535)\pi$.

To check the stability of the solution we added, one by one, new
resonances to the analysis. Adding a resonance with $S_{31}$ or
$G_{37}$ quantum numbers returned masses at the end of the phase
space, very broad widths, negligible fractional contributions  and
marginal improvements in likelihood. For a $D_{35}$ wave, stable fit
results were obtained with $M\approx 2150$\,MeV, $\Gamma\approx
160$\,MeV, and a 1\% fractional contribution. The improvement of the
likelihood was marginal, $\le 10$. Adding a fourth pole to the
$D_{33}$ wave, reproduced the T-matrix pole position reported above
and created an additional pole in the region of 2\,GeV/c$^2$.
However the statistical evidence for this was again marginal,
likelihood changed by $\le 10$. The present data do not support the
need to introduce additional resonances. We tried to improve the
fits by adding decays of $\Delta(2360)D_{33}$ into
$\Delta(1600)P_{33}\eta$ or $N(2200)P_{13}\to N(1710)\eta$. The
statistical significance for these transition was weak, and the
fractional contribution to the data stayed below 0.5\%.

\begin{figure*}[pt]
\epsfig{file=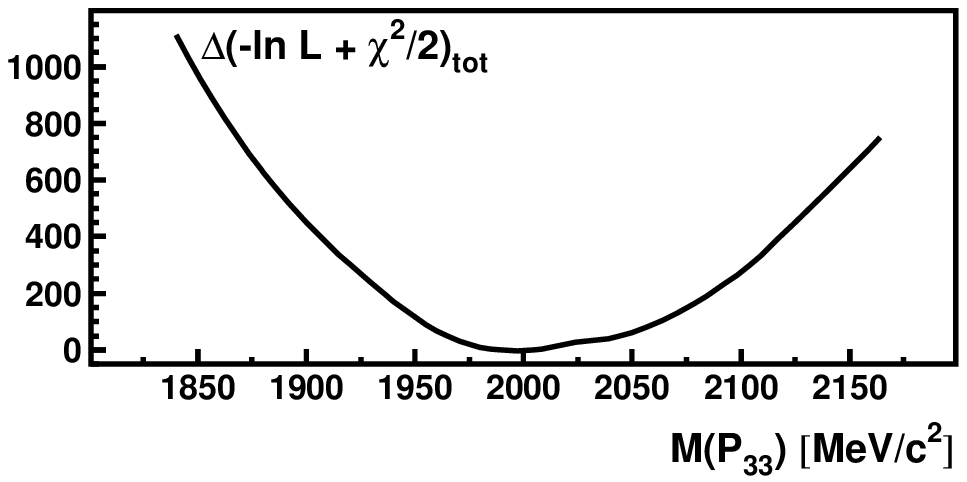,width=0.45\textwidth,height=4.6cm,clip=}
\epsfig{file=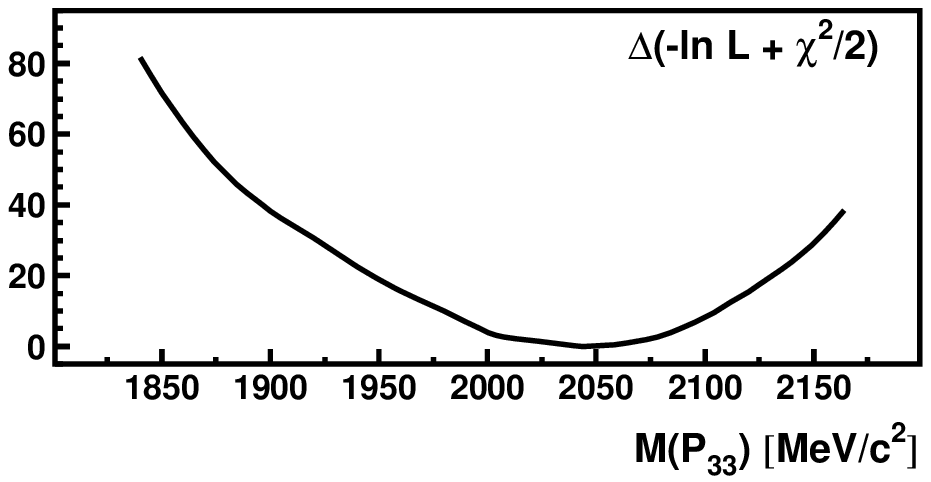,width=0.45\textwidth,height=4.6cm,clip=}
\caption{\label{scanp} Likelihood as a function of the mass of a
$\Delta P_{33}$ resonance. The width is fixed to 350\,MeV/c$^2$. The
left plot gives the change of the total likelihood, the right plot
the change of the likelihood contribution from $\gamma p\to
p\pi^0\eta$, from this data and from beam asymmetry data
\cite{Gutz:2008}. The minimum (reference) value is at about
-280\,000 (a) and -6\,000, respectively. }
\end{figure*}
\begin{figure*}[pt]
\epsfig{file=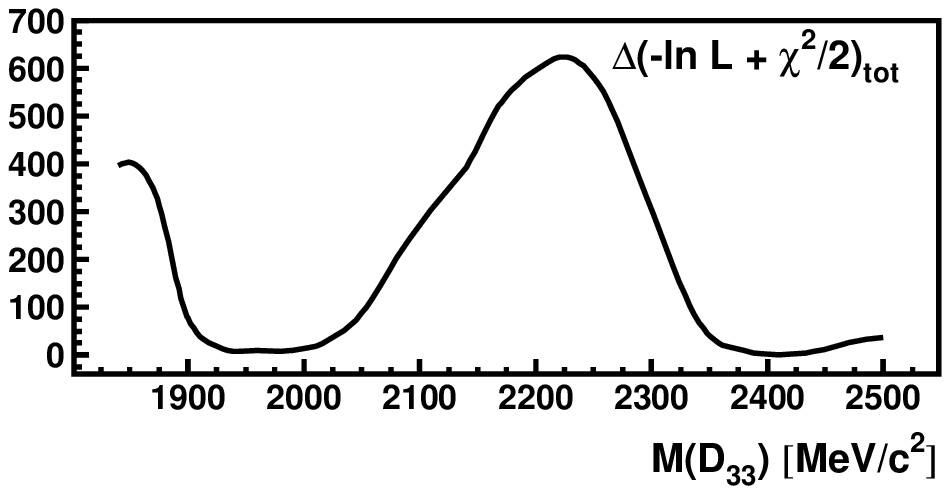,width=0.45\textwidth,height=4.6cm,clip=}
\epsfig{file=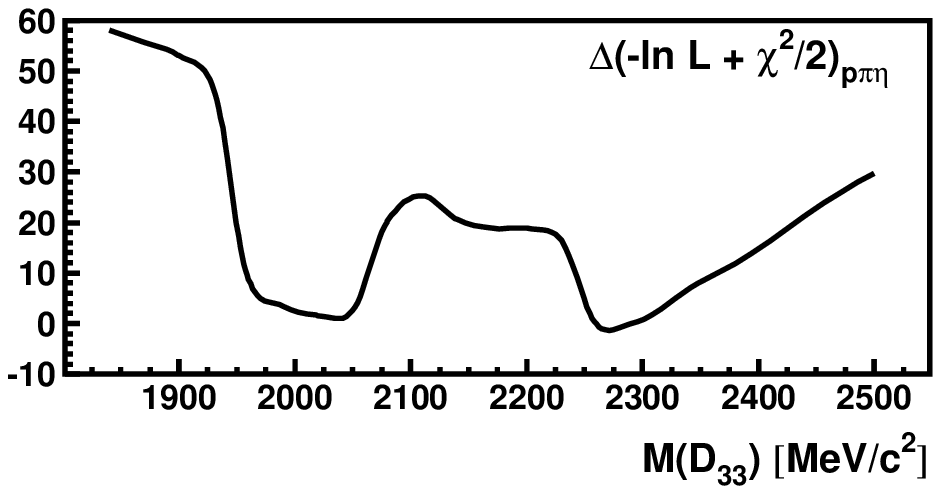,width=0.45\textwidth,height=4.6cm,clip=}
\caption{\label{scand}Likelihood as a function of the mass of a
$\Delta D_{33}$ resonance. The width is fixed to 350\,MeV/c$^2$. The
left plot gives the change of the total likelihood, the right plot
the change of the likelihood contribution from $\gamma p\to
p\pi^0\eta$, this data and beam asymmetry data \cite{Gutz:2008}. }
\end{figure*}

\subsection{Scans}
In a second analysis we replaced, alternatively, the $P_{33}$ or
$D_{33}$ partial wave by a sum of a K-matrix for the energy region
below 1.85\,GeV/c$^2$ and a relativistic Breit-Wigner $(BW)$
amplitude describing the higher-mass region. The mass of the
$P_{33}$ Breit-Wigner state was found to be $1995\pm 50$\,MeV and
the width $380\pm 60$ MeV. The $BW$ mass of the second $D_{33}$
resonance was found to be $1970\pm 50$\,MeV and the width $350\pm
70$\,MeV.

The findings can be visualized in scans showing the changes of $-\ln
\mathcal L_{tot}$ obtained in the fit where the mass of a resonance
under study is changed in steps and fixed while all other variables
are allowed to vary freely. In Fig. \ref{scanp}, the $P_{33}$
amplitude is described by three $BW$ resonances, $\Delta(1232)$,
$\Delta(1600)$ and a third resonance the mass of which is scanned.
The mass of this third resonance was scanned from 1.85 to
2.5\,GeV/c$^2$ and its width fixed to 350\,MeV/c$^2$. Fig.
\ref{scanp} (left) shows the change in total likelihood, Fig.
\ref{scanp} (right) the likelihood chance for the data on $\gamma
p\to p\pi^0\eta$. A clear minimum - which we identify with
$\Delta(1920)P_{33}$ - is observed in both cases; the data on
$\gamma p\to p\pi^0\eta$ have a preference for a somewhat higher
mass.

In Fig. \ref{scand}, a second $D_{33}$ resonance is added to
$\Delta(1700)$. Again, the mass scan covered the 1.85 to
2.5\,GeV/c$^2$ region, and the width was fixed to 350\,MeV/c$^2$.
Wide minima are observed in the 1900-2020\,MeV/c$^2$ region. The
$D_{33}$ wave shows a second minimum at about 2.4\,GeV/c$^2$, in
agreement with the fits described above. The first minimum is
interpreted as evidence for $\Delta(1940)D_{33}$, the second one may
indicate contributions from a higher mass resonance in this partial
wave.

\begin{table*}[pt]
\caption{\label{final}Pole positions, Breit-Wigner parameters, and
decay branching ratios of $\Delta$ resonances. Masses and widths are
given in MeV/c$^2$, branching ratios in \%. The branching ratios are
corrected for Clebsch-Gordan coefficients and unseen decay modes of
final-state mesons ($\pi^0$ and $\eta$). The helicity couplings are
in GeV$^{-1/2}$.\vspace{-2mm}}
\begin{center}
\renewcommand{\arraystretch}{1.5}
\begin{tabular}{ccccccccccc} \hline\hline
&$M_{pole}$&$\Gamma_{pole}$&$M_{BW}$&$\Gamma^{BW}_{tot}$& ${\rm
Br}_{N\pi}$&\hspace{-2mm}${\rm Br}_{\Delta\eta}$\hspace{-2mm}&
\hspace{-2mm}${\rm Br}_{N(1535)\pi}$\hspace{-2mm}&
\hspace{-2mm}${\rm Br}_{Na_0(980)}$\hspace{-2mm}&
\hspace{-2mm}$A_{1/2}$\hspace{-2mm}\hspace{-2mm}&\hspace{-2mm}$A_{3/2}$\\
\hline $\Delta(1600)P_{33}$&$1510^{+20}_{-50}$ & $230\pm 40$&
$1650\pm 40$ & $530\pm 60$
& $10\pm 3$& 0$^{a}$ & 0  & 0  &\multicolumn{2}{c}{not well defined}\\
$\Delta(1920)P_{33}$&$1980^{+25}_{-45}$& $310^{+40}_{-60}$ &
$1990\pm 35$ & $330\pm 60$ & $15\pm 8$ &$ 10\pm5$
& $6\pm 4$ & $4\pm 2$&$22\pm8$&$42\pm12$ \\
$\Delta(1700)D_{33}$& $1640\pm 25$ & $325\pm 35$ & $1790\pm 30$ &
$580\pm 60$& $20\pm 7$
& $2\pm 1$ & $4\pm 2$ & 0&$160\pm40$&$150\pm30$\\
$\Delta(1940)D_{33}$& $1985\pm 30 $ & $ 390\pm 50 $ & $1990\pm 40 $&
$410\pm 70$
& $9\pm 4$ & $4\pm 2$ &$2\pm 1$ & $2\pm 1$\hspace{-2mm}&\hspace{-2mm}$160\pm40$\hspace{-2mm}&\hspace{-2mm}$110\pm30$\\
$\Delta(2360)D_{33}$& $2320\pm 60$ & $500\pm 100$ & $2310\pm 60$ &
$640\pm 120$
& $10\pm 4$ & $60\pm 20$& $7\pm 4$ &$ 9\pm 4$ &$\approx 20$&$\approx 40$\\
\hline\hline
\end{tabular}
\end{center}$^{a}$  The partial width at the Breit-Wigner mass vanishes due to phase
space arguments. The integrated contribution is in the order of a
few \%.
\renewcommand{\arraystretch}{1.0}
\end{table*}

\subsection{PWA results}
Table \ref{final} summarizes our results. The Breit-Wigner masses
and widths and the pole positions are mostly compatible with Review
of Particle Properties (RPP) \cite{Yao:2006px}. Most decay branching
ratios are reported here for the first time. They are calculated as
ratio of partial decay width and total width at the position of the
Breit-Wigner mass. The best known states in Table \ref{final} are
$\Delta(1600)P_{33}$, $\Delta(1700)D_{33}$, and
$\Delta(1920)P_{33}$, all 3-star or 4-star resonances in
\cite{Yao:2006px}. The high rating does however not correspond to a
good agreement in resonance properties derived in different
analyses. In Table \ref{compare} we compare our pole positions with
those of H\"ohler {\it et al.}~\cite{Hohler:1979yr}, Cutkosky {\it
et al.}~\cite{Cutkosky:1980rh}, and Arndt {\it et
al.}~\cite{Arndt:2006bf}. It is of course extremely intriguing that
three out of five of them were not observed in the most recent
analysis~\cite{Arndt:2006bf}. The $\Delta(1600)P_{33}$ $K$-matrix
coupling to $\Delta\eta$ and $N(1535)\pi$ are sizable (see Table
\ref{Table:p33_km}); when the pole is approximated by a Breit-Wigner
resonance, its couplings to these two channels become very small due
to phase space limitations. The K-matrix pole and the Breit-Wigner
mass differ considerably; this fact renders a definition of the
helicity couplings difficult and we refrain from giving numbers.

The $\Delta(1920)P_{33}$ is the third resonance in this partial
wave, after the  well-known $\Delta(1232)$ and the Roper-like
$\Delta(1600)P_{33}$. $\Delta(1920)P_{33}$ was observed in several
analyses including those of H\"ohler {\it et al.}
\cite{Hohler:1979yr} and Cut\-kosky {\it et
al.}~\cite{Cutkosky:1980rh} in elastic $\pi N$ scattering data, and
by Manley {\it et al.} \cite{Manley:1992yb} in a combined analysis
of $\pi N$ elastic scattering and pion induced double-pion
production who found $M$=2057$\pm$110,
$\Gamma$=460$\pm$320\,MeV/c$^2$. In RPP \cite{Yao:2006px} it is
ranked as 3-star resonance. In the analysis by Arndt {\it et al.}
\cite{Arndt:2006bf}, $\Delta(1920)P_{33}$ was not found.
$\Delta(1600)P_{33}$ is observed in all four analyses of Table
\ref{compare}; in \cite{Sarantsev:2005tg}, it was not needed to get
a good description of the data taken into account at that time.

The  $\Delta(1940)D_{33}$ state has only one star in the RPP
classification. It was observed by Cut\-kosky {\it et} {\it
al.}~\cite{Cutkosky:1980rh} in elastic $\pi N$ scattering data with
mass $M$=$1940\pm100$\,MeV/c$^2$ and width $\Gamma=
200\pm100$\,MeV/c$^2$,~and by Chew {\it et al.} \cite{Chew:1980jn}
at $M$=$2058\pm35$\,MeV/c$^2$ and $\Gamma$=198$\pm$46\,MeV/c$^2$
while Manley found $M$=$2071\pm100$\,MeV/c$^2$,
$\Gamma$=$460\pm320$\,MeV/c$^2$ \cite{Manley:1992yb}. In
\cite{Arndt:2006bf}, $\Delta(1940)D_{33}$ was not found.

\begin{table}[pt]
\caption{\label{compare}Pole positions (in MeV/c$^2$) of ``well
established" $\Delta$ resonances and the one-star
$\Delta(1940)D_{33}$.} \begin{center}
\renewcommand{\arraystretch}{1.4}
\begin{tabular}{lcccc} \hline\hline
\hspace{-2mm}Resonance \hspace{-3mm} &\hspace{-3mm}H\"ohler\hspace{-3mm}&\hspace{-3mm}Cutkosky\hspace{-3mm}& Arndt  &\hspace{-3mm}this work\hspace{-3mm}\\
\hline
\hspace{-2mm}$\Delta(1600)P_{33}$&\hspace{-13mm}$1550$\hspace{-4mm}&\hspace{-4mm}$1550-i100$\hspace{-3mm}&\hspace{-3mm}$1457-i200$\hspace{-3mm}&\hspace{-3mm}$1510-i115$\hspace{-3mm}\\
\hspace{-2mm}$\Delta(1700)D_{33}$&\hspace{-3mm}$1651-i80$\hspace{-4mm}&\hspace{-4mm}$1675-i110$\hspace{-3mm}&\hspace{-3mm}$1632-i126$\hspace{-3mm}&\hspace{-3mm}$1650-i160$\hspace{-3mm}\\
\hspace{-2mm}$\Delta(1920)P_{33}$&\hspace{-13mm}$1900$\hspace{-4mm}&\hspace{-4mm}$1900-i150$\hspace{-3mm}&\hspace{-3mm}not~seen\hspace{-3mm}&\hspace{-3mm}$1980-i155$\hspace{-3mm}\\
\hspace{-2mm}$\Delta(1905)F_{35}$&\hspace{-3mm}$1829-i152$\hspace{-4mm}&\hspace{-4mm}$1830-i140$\hspace{-3mm}&\hspace{-3mm}not~seen\hspace{-3mm}&\hspace{-3mm}$1920-i145$\hspace{-3mm}\\
\hspace{-2mm}$\Delta(1940)D_{33}$&\hspace{-5mm}not seen\hspace{-4mm}&\hspace{-4mm}$1900-i100$\hspace{-3mm}&\hspace{-3mm}not~seen\hspace{-3mm}&\hspace{-3mm}$1985-i195$\hspace{-3mm}\\
\hline\hline
\end{tabular}
\renewcommand{\arraystretch}{1.0}
\end{center}
\vspace{-4mm}
\end{table}
The confirmation of the two resonances $\Delta(1920)P_{33}$ and
$\Delta(1940)D_{33}$ in a new reaction is a major result of this
analysis. In addition, several partial decay branching ratios have
been determined. These provide first insight into an old question on
how the mass of high-lying parent resonances is dissipated, into
high-momenta of the daughter particles, into high-mass mesons or
into high-mass daughter baryon resonances. The results suggest that
high-mass parent baryons seem to take advantage of all allowed decay
modes, including intermediate meson {\it and} baryon resonances even
at rather high masses. The situation resembles $\bar NN$
annihilation \cite{Vandermeulen:1988hh,Klempt:2005pp}: in this
process, the production of high-mass mesons is preferred in
comparison to production of low-mass mesons even though the phase
space is of course larger for the latter reaction. Likewise, there
seems to be no preference for the kinematically favored $N\pi$ decay
mode.

\section{Summary and conclusions\vspace{2mm}}
We have reported a study of the photoproduction process $\gamma p\to
p\pi^0\eta$ at the Electron Stretcher Apparatus ELSA at Bonn. About
16\,000 $p\pi^0\eta$ events were extracted above a small background.
For small photon energies, the cross section for the reaction agrees
with previous results from Sendai and Graal. In this energy region,
the process is dominated by the $\Delta(1232)\eta$. At higher photon
energies, a significant fraction of the process proceeds via the
$N(1535)\eta$ and some $p\,a_0(980)$ contribution is observed.

A partial wave analysis determines the resonant contributions. Two
waves with $P_{33}$ and $D_{33}$ quantum numbers, with $J^P=3/2^+$
and $3/2^-$, dominate the reaction. We find evidence for two
resonances $\Delta(1920)P_{33}$ and $\Delta(1940)D_{33}$ and
indications for a further high-mass $D_{33}$ resonance. The evidence
for the two states was already communicated in a letter publication
\cite{Horn:2008a}.

A detailed discussion of the role of $\Delta(1920)P_{33}$ and
$\Delta(1940)D_{33}$ within the spectrum of $\Delta$ resonances will
be given elsewhere \cite{Forkle}. Here we just mention the main
results.
\begin{enumerate}
\item In \cite{Forkle}, the 3 negative parity resonances $\Delta(1900)S_{31}$,
$\Delta(1940)D_{33}$, and $\Delta(1930)D_{35}$ are interpreted as
triplet of resonances with $L=1, S=3/2, N=1$ coupling to
$J=1/2,\,3/2,\,5/2$ and belonging to the $(D,L^P_{\rm N})\,S =
(56,1^-_3)\,3/2$ supermultiplet. Alternatively, $\Delta(1900)S_{31}$
and $\Delta(1940)D_{33}$ could be a $L=1, S=1/2, N=1$ doublet, and
$\Delta(1930)D_{35}$ and $\Delta(2200)G_{37}$ a $L=3, S=1/2$
doublet. The mass of the triplet is rather low compared to quark
model predictions \cite{Capstick:1986bm,Loring:2001kx} while the
$\Delta(2200)G_{37}$ mass is perfectly compatible with the
models.\vspace{2mm}
\item The two resonances $\Delta(1920)P_{33}$ and
$\Delta(1940)D_{33}$ form a parity doublet \cite{Glozman:1999tk}.
Parity doublets are of topical interests, we refer the reader to two
recent reviews \cite{Jaffe:2006jy,Glozman:2007jt}. In the latter
review, there is a written request to search for the doublet
reported here. The conjecture that the reason for the mass
degeneracy is due to a restoration of chiral symmetry in high-mass
baryon states falls of course outside of experimental observability,
and the interpretation is controversially discussed
\cite{Klempt:2002tt,Klempt:2007cp,Shifman:2007xn}.\vspace{2mm}
\item All masses in the $\Delta$ spectrum are perfectly compatible
with models based on AdS/QCD when confinement is modeled by a soft
wall in the holographic variable \cite{Forkel:2007cm}. The hard wall
approximation \cite{Brodsky:2006uq} gives a good general survey but
fails in important details.
\end{enumerate}

\section*{Acknowledgement}
We would like to thank the technical staff of ELSA and of the
participating institutions for their invaluable contributions to the
success of the experiment. We acknowledge financial support from the
Deutsche Forschungsgemeinschaft (DFG-TR16) and the Schweizerische
Nationalfond. The collaboration with St. Petersburg received funds
from DFG and RFBR. This work comprises part of the thesis of I.
Horn.

\end{document}